\documentclass[a4paper,11pt]{article}
\pdfoutput=1 % if your are submitting a pdflatex (i.e. if you have
\usepackage{jcappub} % for details on the use of the package, please
\usepackage[T1]{fontenc} % if needed

\usepackage{xcolor}

\usepackage{caption, subcaption}

\usepackage{graphicx}

\usepackage{amsmath}

\usepackage{footmisc}

\usepackage{booktabs,siunitx}

\usepackage[toc]{appendix}

\title{\boldmath Primordial Power Spectrum Reconstruction From CMB Weak Lensing Power Spectrum}

\author[a,1]{Rajorshi Sushovan Chandra,\note{Corresponding author.}}
\author[a,b]{and Tarun Souradeep}

\affiliation[a]{Inter-University Centre for Astronomy and Astrophysics, Post Bag 4, Ganeshkhind, \\Pune 411 007, India}
\affiliation[b]{Indian Institute of Science Education and Research, Dr. Homi Bhabha Road, Pashan, \\Pune, 411 008, India}
\abstract{We use the modified Richardson-Lucy deconvolution algorithm to reconstruct the Primordial Power Spectrum from the Weak Lensing Power spectrum reconstructed from the CMB anisotropies. This provides an independent window to observe and constrain the PPS $P_R(k)$ along different $k$ scales as compared to CMB Temperature Power Spectrum. The Weak Lensing Power spectrum does not contain secondary variations in power and hence is cleaner, unlike the Temperature Power spectrum which suffers from lensing which is visible in its PPS reconstructions. We demonstrate that the physical behaviour of the weak lensing kernel is different from the temperature kernel and reconstructs broad features over $k$. We provide an in-depth analysis of the error propagation using simulated data and Monte-Carlo sampling, based on Planck best-fit cosmological parameters to simulate the data and cosmic variance limited error bars. The error and initial condition analysis provides a clear picture of the optimal reconstruction region for the estimator and we provide and algorithm for $P_R(k)$ sampling to be used based on the given data, errors and its binning properties. Eventually we plan to use this method on actual mission data and provide a cross reference to PPS reconstructed from other sectors and any possible features in them. }

\begin{document}
\maketitle
\flushbottom

\section{Introduction}
\label{sec:intro}

The $\Lambda$CDM Standard Model of cosmology has been a mainstay in the field of cosmology for a long duration and is now in the realm of precision cosmology. Cosmic Microwave Background anisotropies have given us a powerful natural laboratory to infer the parametrizations used in $\Lambda$CDM. The Planck \cite{Aghanim:2018eyx} experiment and collaboration has provided a precise and consistent analysis of the data and parameter inference. A section of this analysis involves a parametrization of the Primordial Power Spectrum (PPS), which is usually modelled as a power law with the two parameters, amplitude $A_s$ and power index $n_s$. This is by no means the last word as many other parametrizations exist, from light adjustments like a $k$ dependent $n_s$ (running the power law), to specific model dependent features in the PPS like steps and valleys. Given the large and diverse range of inflation models, there is a well established framework that seeks to reconstruct the PPS using a model independent approach, which essentially treats each PPS ($k$) sample as a free parameter. Much work has been done on this approach, using the Richardson-Lucy deconvolution algorithm \cite{Shafieloo:2003gf,Shafieloo:2007tk,Hazra:2013xva,Hazra:2014jwa}. The basic working premise of this algorithm is in the mathematical property that the power spectra $C_L,P_R{(k)}$ and transfer functions $G_{Lk}$ are positive definite, being squares of the underlying Gaussian random fields and transfer functions. It has been successfully implemented on the temperature and polarisation power spectra from CMB anisotropies. 
Recent precision reconstructions of the weak lensing power spectrum $C_L^{\phi\phi}$ from Planck, SPT, ACT \cite{Wu:2019hek,Millea:2020iuw,Sherwin:2016tyf,Das:2013zf,2011PhRvL.107b1301D} and several other surveys, based on state of the art reconstruction formalisms, have opened up the possibility of reconstruction the PPS from this data. Since the physical processes encoded in the transfer function of weak lensing is different from that of the CMB anisotropy transfer function, it is of great interest to study this sector as way to cross validate PPS reconstructions from other sectors. In recent years there has also been renewed interest in seeking alternatives to the established power law paradigm in order to find a resolution to interesting discrepancies in data, such as the $H_0$ tension \cite{Hazra:2018opk,Keeley:2020rmo}. Also, further improvements are expected in the weak lensing power spectrum $C_L^{\phi\phi}$ observations from the S4 experiments and a future full-sky mission. In light of these, it is a worthwhile exercise to observe the behaviour of the MRL algorithm on the weak lensing transfer kernel and the statistical properties of a PPS reconstruction that can be obtained form $C_L^{\phi\phi}$. 

In this paper we will first give an overview of the weak lensing transfer functions and some numerical details of calculating it in section \ref{sec:wklensclpp}. Then we give a brief overview of the MRL algorithm and its applicability on $C_L^{\phi\phi}$ in section \ref{sec:IRLdeconv}. After that we simulate the transfer kernel and provide a detailed view of its physical properties and transfer behaviour from PPS to $C_L^{\phi\phi}$ as well as reconstruction features, limitations and initial guess sensitivity in section \ref{sec:kernelsim}. After that in section \ref{sec:ressimtest} we simulate $C_L^{\phi\phi}$ under specific conditions and error budgets and carry out a reconstruction of the PPS, showing the effect of our new $P_R{k}$ sampling algorithm that improves statistical significance of reconstruction. Finally we give a detailed overview of the statistical significance of our results and the overall robustness of the MRL algorithm in the $C_L^{\phi\phi}$ sector, for future reference for application to actual data.

\section{Weak Lensing Angular Power Spectrum $C_L^{\phi\phi}$}
\label{sec:wklensclpp}

The weak lensing power spectrum forms a distinct measure of the cosmological parameters given the baseline $\Lambda$CDM model with non-relativistic cold dark matter and a spatially flat FLRW cosmological model. %(These aspects will also be briefly addressed post analysis of the data). 
Due to this, we have an independent probe of the cosmological model information by observing the $C_L^{\phi\phi}$ reconstructed from CMB observations. In this context a deconvolution of the primordial power spectrum $P_{R}(k)$ from the lensing power spectrum given by 

\begin{equation}
\begin{split}
C_L^{\phi\phi} = \sum_{k} G_{Lk} P_{R}(k)
\end{split}
\end{equation}

\noindent can provide us with an independent probe to reconstruct the primordial power spectrum as the primordial curvature perturbations have evolved via a different sector of physical effects than CMB, resulting in an independent observable power spectra, encoded in the CMB via lensing. We provide a brief overview of the physics of weak lensing and its relation to the primordial comoving curvature perturbations. The weak lensing potential map and its power spectrum are reconstructed from CMB, which is observed as a lensed CMB anisotropy map, where the weak lensing effect is expressed by 

\begin{equation}
\tilde{\Theta}(\hat{n}) = \Theta(\hat{n}+\vec{\alpha} )
\end{equation}

\noindent This remapping angle $\vec{\alpha}$ is in turn related to the weak lensing potential of the cosmological matter distribution and its effects integrated over all redshifts

\begin{equation}
\vec{\alpha} = -2 \int_{0}^{\chi_*} 
\nabla_{\hat{n}}\Phi(\chi\hat{n};\eta_0-\chi) \frac{f_K(\chi_*-\chi)}{f_K(\chi_*)f_K(\chi)} d\chi
\end{equation}

\noindent  Where the light from the source plane (last scattering surface) $\chi_*$ is being lensed by an instance of a lensing potential $\Phi$ at a comoving distance $\chi$ in the direction of the line of sight $\hat{n}$. This is then integrated along all redshifts. The  cumulative effect is expressed as the remapping angle $\alpha$ due to the cumulative CMB lensing potential projected over the sky. 
The remapping angle can be expressed as the gradient of this projected lensing potential

\begin{equation}
\begin{split}
& \vec{\alpha} = \nabla_{\hat{n}}\phi(\hat{n}) \\
& \phi(\hat{n}) = -2 \int_{0}^{\chi_*} 
\Phi(\chi\hat{n};\eta_0-\chi) \frac{f_K(\chi_*-\chi)}{f_K(\chi_*)f_K(\chi)} d\chi
\end{split}
\end{equation}

\noindent The lensing potential instance $\Phi(\chi\hat{n};\eta_0-\chi)$ at any given comoving distance can be expressed as the evolution of the primordial curvature perturbation $R(k)$ by a transfer function, which can be expressed as

\begin{equation}
\begin{split}
\Phi(\chi\hat{n};\eta_0-\chi) =  T_{\Phi}(k;\eta) R(k)
\end{split}
\end{equation}

\noindent  We have used a cosmological model which assumes linear evolution of the primordial Gaussian fields and hence we are discounting any non linear effects on the $C_L^{\phi\phi}$ (which will have significant contribution from non linear effects at small scales of about $L \approx 2000$ and beyond) when we model the transfer kernel $T_{\Phi}$.
The transfer equation for the lensing potential power spectrum, in the linear regime is given by 

\begin{equation}
C_L^{\phi\phi} = 4\pi \int P_{R}(k) \bigg[\int_0^{\chi_{*}} 2T_{\phi}(k;\eta_0-\chi)\bigg(\frac{(\chi_{*}-\chi)}{\chi_*\chi}\bigg)j_L(k\chi) d\chi \bigg]^2 \frac{dk}{k}
\label{eq:clppexact}
\end{equation}

\noindent Our observable is the lensing power spectrum reconstructed from the lensed temperature map, given by

\begin{equation}
\begin{split}
C_{L}^{\kappa\kappa} &= \frac{[L(L+1)]^2}{2\pi} C_L^{\phi\phi} \\
&= \frac{[L(L+1)]^2}{2\pi} 4\pi \int P_{R}(k) \bigg[\int_0^{\chi_{*}} 2T_{\phi}(k;\eta_0-\chi)\bigg(\frac{(\chi_{*}-\chi)}{\chi_*\chi}\bigg)j_L(k\chi) d\chi \bigg]^2 \frac{dk}{k}
\end{split}
\end{equation}

%2) We now have the $G\bigg(k,z(\frac{(L+\frac{1}{2})}{k})\bigg)$ for a given $L$ and $\forall k \in [k_*(L) ... k_{max}]$

\noindent In the Limber approximation (Accurate at high L) case it is

\begin{equation}
C_{L}^{\kappa\kappa} = 2{[L(L+1)]^2}  \int P_{R}(k) \bigg[ \sqrt{\frac{\pi}{2L}} \frac{1}{k} 2T_{\phi}(k;\eta_0-\chi_s)\bigg(\frac{\chi_{*}-\chi_s}{\chi_*\chi_s}\bigg) \bigg]^2 \frac{dk}{k} \  \forall \ \chi_s = \frac{L}{k}
\label{eq:clpplimber}
\end{equation}

\noindent For our purposes we will simulate the transfer kernel of the power spectrum by the exact expression for $L \in [2 \rightarrow 100]$ and the Limber approximated expression for $L > 100$ to increase computation efficiency, while retaining numerical accuracy.

\noindent So our numerical expression for the kernel using the Trapezoidal rule, for the exact expression, we have

\begin{equation}
\begin{split}
C_{L}^{\kappa\kappa} &\approx 
8{[L(L+1)]^2} \sum_{k=k_{min}}^{k_{max}} P_{R}(k) 
\bigg[ \sum_{1}^{\chi_*} T_{\phi}^2(k;z(\chi)) 
\mathcal{W}^2(\chi)
j_L^2(k\chi) \Delta\chi \bigg] 
\frac{\Delta k}{k} \\
\text{ Where } \mathcal{W}(\chi) &= \bigg(\frac{\chi_{*}-\chi}{\chi_*\chi}\bigg)
\end{split}
\end{equation}

\noindent For the Limber approximated expression we have

\begin{equation}
\begin{split}
C_{L}^{\kappa\kappa} &= 4\pi\frac{[L(L+1)]^2}{L}  
\sum_{k=k_*(L)}^{k_{max}} P_{R}(k) 
\bigg[ T^2_{\phi}(k;z(\chi_s))
\bigg(\frac{k}{L+0.5}-\frac{1}{\chi_*}\bigg)^2 \bigg]
\frac{\Delta k}{k^3} \\
\chi_s &= \frac{L+0.5}{k} \ \ \ \ 
k_*(L)  = \frac{(L+0.5)}{\chi_*}
\label{eq:clpplimber}
\end{split}
\end{equation}

\iffalse
\begin{equation}
\begin{split}
C_l^{\phi\phi} &= \frac{4\pi}{l} \int \ 2\pi \ P_{R}(k)  \bigg[ G\bigg(k,z(\frac{(l+\frac{1}{2})}{k})\bigg) \bigg]^2 \bigg(\frac{k}{l} - \frac{1}{\chi_*}\bigg)^2  \frac{dk}{k^3} \\
C_l^{\phi\phi} &\approx \frac{8\pi^2}{l} \sum_{k = k_*(l)}^{k_{max}}= P_{R}(k)  \bigg[ G\bigg(k,z(\frac{(l+\frac{1}{2})}{k})\bigg) \bigg]^2 \bigg(\frac{k}{l+0.5} - \frac{1}{\chi_*}\bigg)^2  \frac{\Delta k}{k^3} \ \text{where} \ k_*(l)  = \frac{(l+0.5)}{\chi_*}
\end{split}
\end{equation}
\fi

%\begin{equation}
%C_{L}^{\kappa\kappa} = \frac{[L(L+1)]^2}{2\pi} C_{L}^{\phi\phi}
%\end{equation}

So we can put the net expression in the operational form of a discrete convolution 

\begin{equation}
C_{L}^{\kappa\kappa} = \sum_{k} G_{Lk} P_R(k)
\end{equation}

%\textcolor{blue}{1. Denote difference between GLk and GL(k). Write detailed kernel equations and make notation consistent.} 
\noindent Here we use the following notation for the kernel for the discrete form to be used in the numerical calculation in the R-L estimator and the functional form

\begin{equation}
\begin{split}
G_{Lk} &= G_L(k)\Delta k \\
G_L(k) &= 
\begin{cases} 
      \frac{8{[L(L+1)]^2}}{k}
\bigg[ \sum_{1}^{\chi_*} T_{\phi}^2(k;z(\chi)) 
\mathcal{W}^2(\chi)
j_L^2(k\chi) \Delta\chi \bigg] 
& 2 \leq L \leq 100,\ k_{min} \leq k \leq k_{max} \\
      4\pi\frac{[L(L+1)]^2}{Lk^3}
\bigg[ T^2_{\phi}(k;z(\chi_s))
\bigg(\frac{k}{L+0.5}-\frac{1}{\chi_*}\bigg)^2 \bigg] 
& 100 < L,\ k_{*}(L) \leq k \leq k_{max} \\
\end{cases}
\end{split}
\end{equation}

\noindent This expression can be computed by obtaining the Transfer function using the public CAMB software for a specified chosen cosmological model \cite{Lewis:1999bs}.

%\textcolor{red}{1. The actual kernel used in estimation will be the term $G_L(k)\Delta k$ as this is the area under the convolution being carried out and its visualisation in the graphs give an understanding of how power transfer is distributed along different scales. } 

Given this theoretical model we can observe that the observable power spectra is indeed an independent estimator of the primordial curvature perturbation power spectrum, which is unmodified by any significant higher order physical process that is also dependent on the $P_R(k)$. We recall the weak lensing of the CMB temperature power spectrum where the lensed contribution is itself a function of $P_R(k)$, making it a non linear dependency on the PPS and necessitating a template based 'de-lensing' approach before deconvolution. In the following section we will explain the deconvolution process followed and the technical details involved.
In section \ref{sec:kernelsim}, Results Ia, we provide the details of the kernel simulation.

\section{Method : Improved Richardson-Lucy Deconvolution}
\label{sec:IRLdeconv}

%\textcolor{blue}{2. Put all the math of the IRL algorithm}	\\

To proceed with the deconvolution of the primordial curvature perturbation power spectrum $P_R(k)$ convolved by the radiative transfer function into the observable $C_{L}^{\kappa\kappa} $

\begin{equation}
C_{L}^{\kappa\kappa} =  \sum_{k} G_{Lk} P_R(k)
\end{equation}

We use a deconvolution algorithm called the Richardson-Lucy deconvolution. \cite{1972JOSA...62...55R, 1974AJ.....79..745L} While the detailed derivation and its relation to other standard estimators may be looked up in previous literature \cite{Shafieloo:2003gf,Shafieloo:2007tk,Nicholson:2009pi,Hazra:2013xva,Hazra:2014jwa}, in principle it works by treating the transfer function $G_{Lk}$ as a probability distribution over the $L$ index (Hence the normalisation to unity over L). The estimator iteratively adds the 'remnant power' weighted by the probability distribution $\tilde{G}_{Lk}$ to the previous iteration of $P_k^{(i)}$, starting from some guess $P_k^{(0)}$. The estimator is valid for positive definite functions being convolved, which is applicable here as we are working with the power spectra of Gaussian random fields in linear regimes.
 
%\textcolor{blue}{7. What should be the stopping criteria ? \footnotemark[1]\footnotetext[1]{ Hazra and Shaf use the chisqstat of the reconstructed CL using the likelihood function of Planck/WMAP. There the goal was to show that a freeform PRk recon CL fits the data better than a powlaw PRk recon CL and that once IRL freeform PRk does its best, the chisqstat won't improve anymore and can stop reconstruction. Contaldi et al use a simpler version where iteration stops once $\Delta$PRk is less than 0.1\%.	\\
%\textcolor{red}{Use $\Delta C_C^{\phi\phi}$ 0.1\% convergence test since it is robust and converges fast. reasoning is that it is more sensitive at all L rather than Pk which does not converge ate low k}
%We are using the Shaikh method where the IRL iteration stops when the reconstructed $\Delta C_L^{(i)}$ with respect to previous iteration reaches a saturation at 0.1\%. This is a quick and consistent way to reach convergence, even though there exist other more expensive methods and given the existing literature and their results we shall use this technique.} }

We need a stable and accurate stopping criteria. While there are several options in the existing literature, such as minimizing the $\chi^2$ of the reconstructed $C_{L}^{\kappa\kappa}$ with respect to the WMAP or Planck likelihood or comparing the change in the -ln$\mathcal{L}$ with respect to number of iterations \cite{Shafieloo:2003gf}. Other methods include simply checking for a 0.1\% relative error cap between the final two iterations \cite{Nicholson:2009pi} However for our work, we compare the relative difference between the final two iterations of the reconstructed $C_{L}^{\kappa\kappa}$ across all multipoles and stopping iteration when at all multipoles are below 0.1\% relative difference. The reconstructed $P_R(k)$ does not necessarily converge to a stable reconstruction at multipoles where the propagated error bars are the largest. However seeing as it is a poorly constrained set of equations, this is to be expected because of degeneracy in solutions. However the reconstructed $C_{L}^{\kappa\kappa}$ is much more stable with respect to iterations and hence we use it to identify the iteration stopping point.

In addition to the original R-L estimator, previous literature \cite{Shafieloo:2003gf} have provided a method to make it sensitive to the error in the data, since it is a fundamental feature of the observed power spectrum, being limited by cosmic variance even in an ideal scenario. To account for this the R-L estimator is updated to the IRL which weighs each multipole $L$ by its relevant error bar and correlations between them.

\begin{equation}
\begin{split}
P_{k}^{(i+1)} &= P_{k}^{(i)} \bigg{[} 1 + 
\sum_{L} \tilde{G}_{Lk} \bigg{(} \frac{\hat{C}_{L}^{\kappa\kappa}}{C_{L}^{\kappa\kappa(i)}} - 1 \bigg{)}
\text{tanh}^2 \big{(} 
[\hat{C}_{L}^{\kappa\kappa} - {C}_{L}^{\kappa\kappa(i)}]
\Sigma^{-1}
[\hat{C}_{L}^{\kappa\kappa} - {C}_{L}^{\kappa\kappa(i)}]^T 
\big{)} \bigg{]} \\
&= P_{k}^{(i)} \bigg{[} 1 + 
\sum_{L} \tilde{G}_{Lk} \bigg{(} \frac{\hat{C}	_{L}^{\kappa\kappa}}{C_{L}^{\kappa\kappa(i)}} - 1 \bigg{)}
\text{tanh}^2 \bigg{(} \frac{\hat{C}_{L}^{\kappa\kappa}-C_{L}^{\kappa\kappa(i)}}{\hat{\sigma}_{L}} \bigg{)}^2 \bigg{]} \\
\tilde{G}_{Lk} &\text{ : Discretized Kernel normalised over L} \\
\hat{C}_{L}^{\kappa\kappa} &\text{ : Data $C_{L}^{\kappa\kappa}$} \\
\Sigma^{-1} &: \text{Error covariance matric of the data}\ \hat{C}_{L}^{\kappa\kappa} \\
C_{L}^{\kappa\kappa(i)} &=  \sum_{k} G_{Lk} P^{(i)}_k \\
\end{split}
\label{eq:IRL_eqns}
\end{equation}

%\textcolor{blue}{3. Free form means each P of k is a free variable}

%\textcolor{blue}{4. Is it good that since GLKPP is smooth over k, that means fewer Pk as variable, hence faster calculations, less degeneracy and correlation ? (At the cost of losing high frequency features} 

One should note that this estimator finds a free form estimate of the $P_R(k)$. What this means in principle is that each $P_R(k_i)$ is a variable being estimated. In application this means that the number of unknowns will depend on the number of gridpoints over $k$ used to evaluate $P_R(k)$. A large number of gridpoints increase the possibility of finding sharp features over the $k$ range, but increase the correlation between different $P_R(k_i)$ assuming the number of data points over $L$ remain the same. In effect, one can only optimize what aspect of the data to infer best given a finite amount of data. In this regard it will be seen that the transfer kernel $G_{Lk}$ is relatively smooth over $k$ and damps out any sharp features in $P_R(k)$ during convolution, leading to degeneracy between features that may contribute to the observable $\hat{C}_{L}^{\kappa\kappa}$. The conclusions from this observation are that we can reduce the number of gridpoints over $k$ at which to evaluate the different $P_R(k_i)$. We expect this approach to reduce the correlation between different $P_R(k_i)$ and reduce computation time, but also restrict the sensitivity of the estimator to sharp/high frequency features in $P_R(k)$. We highlight that this restriction is a fundamental feature of the physics present in the transport kernel $G_{Lk}$ and only broad features over $P_R(k)$ can be resolved given this estimator. One simple way to mitigate this limitation is to use independent cross checks with other power spectra such as the $\hat{C}_{L}^{TT}$ whose transport kernel $G_{Lk}^{TT}$ have different sensitivities to different $k$ ranges and can discover features in $P_R(k)$ that $G_{Lk}^{\kappa\kappa}$ will miss.
\\

%\footnotemark[5]\footnotetext[5]{
%\textcolor{red}{5. Mention the understanding that calculating the chisuquare statistic of the recovered set of free from PofKs means testing whether Pks follow a Gaussian disstribution based on the value of chisquare stat and number of Pks; NPk. This means that even though theoretically both CL and Pk are chi square distributed variables because the $\phi$ and $\psi$ fields are gaussian random fields, we can assume they are Gaussian distributed (especially considering that at high L, the number of averaged squares of gaussian, azimuthal modes are 2L+1 and they converge to Gaussian for high L.}

%\footnotemark[5]\footnotetext[6]{
%\textcolor{red}{6. Mention that there is faint correlation between the Pks and hence it might be a nice exercise eventually to diagonalise the cov matrix of Pk and find the rotation matrix, apply that to the GLK kernel and CL data so that we recover the PD(k) in the diagonal basis where each parameter PD is uncorrelated to the other PD, where PD is a linear combination of the P's. However, stress that this is beyond the scope of this paper and can be reserved for future work.}

\section{Results Ia : Kernel Simulation}
\label{sec:kernelsim}

\subsection{Kernel Behaviour}
\label{sub_sec:kernel_behav}

In this section we will discuss the behaviour of the simulated radiative transport kernel that we will use to deconvolve the primordial power spectrum from the observed angular power spectrum data realisation. We have used the Planck \cite{Aghanim:2018eyx} inferred best fit background cosmological parameters as given in Table \ref{table:phy_param} 
%\textcolor{red}{Cite[Planck Param Paper]}
%\textcolor{red}{Cite[Self Table]}.

To simulate the $G_{LK}^{\phi\phi}$ kernel we used the CAMB \cite{Lewis:1999bs} software. The numerical parameters used to simulate the kernel are given in Table \ref{tab:sim_param}
% \textcolor{red}{Cite[CAMB, Lewis]}

%We provide a justification for limiting ourselves to the $k_{max}=1$. 
%Given a fiducial power spectrum $C_L^{\phi\phi}$, our fiducial model must be accurate to within the Cosmic Variance precision when applied to data. If we lower the cutoff $k_{max}=1$, there will be a corresponding loss of power visible as a bias which can affect our reconstruction. 
%We vary the $k_{max}$ and see that below a $k_{max}=1$, the loss of power bias exceeds cosmic variance, which we have set as the min error in observed data upto an $L_{max}=2500$, Figure \ref{fig:kmaxcut_a}.
%Hence under these $k$ and $L$ limits, we restrict our reconstruction range to $k \in [10^{-6}, 1]$.

\begin{table}
\emph{\begin{center}
%s\noindent
\begin{tabular}{ l r r l }
  \toprule
  Simulation Parameters & \multicolumn{1}{c}{Simulation Input} \\
  \midrule
  $k_{min}$ & \num{7E-6} \\
  $k_{max}$ & \num{1E+1 } \\
  $N_{k}$ & \num{3151 } \\
  $L_{min}$ & \num{2 } \\
  $L_{max}$ & \num{2500 } \\
  $\Delta \log{k}$ & \num{500 } \\
  $Simulation Accuracy$ & \num{8 } \\
  \bottomrule
\end{tabular}
\end{center}}
\caption{The simulation parameters with reference to CAMB. The $k$ range choice is explained in the text. The $L$ range choice is based on currently observational data. The number of free variables $N_{k}$ is automatically assigned on the basis of numerical parameters such as \noindent $\Delta\log{k}$ (the number of $k$ intervals on the log $k$ scale) and a CAMB specific simulation accuracy parameter that controls $k$ sample resolution called $SourcekAccuracyBoost$. %\textcolor{red}{$SourcekAccuracyBoost$ parameter decision was based on a rough understanding of CAMB documentation and trial and error. There may be some room to understand CAMB more deeply, but for the purposes of this paper it under control.}
}
\label{tab:sim_param}
\end{table}

%\begin{center}
%s\noindent
%\begin{tabular}{ l r r l }
 % \toprule
%  Dataset Filename & \multicolumn{1}{c}{Cases} & \multicolumn{1}{c}{Records} & Primary Key \\
 % \midrule
 % Detail Interaction & \num{147004} & \num{2400} & InteractionID \\
 % Detail Incident & \num{12345} & \num{20000} & IncidentID \\
 % Detail Change & \num{12} & \num{412} & ChangeID \\
 % Detail Activity & \num{7890} & \num{1234567} & ActivityID \\
 % \bottomrule
%\end{tabular}
%\end{center}

%\begin{table}
%\emph{\begin{center}
%%s\noindent
%\begin{tabular}{ l r r l }
%  \toprule
%  Parameter & \multicolumn{1}{c}{Planck Best-Fit} \\
%  \midrule
%  $H_0$ & \num{0.6732117E+02} \\
%  $\Omega_b h^2$ & \num{0.2238280E-01 } \\
%  $\Omega_c h^2$ & \num{0.1201075E+00 } \\
%  $m_\nu$ & \num{0.6000000E-01 } \\
%  $\Omega_K$ & \num{0} \\
%  $\tau$ & \num{0.5430842E-01} \\
%  $A_s$ & \num{2.100549E-09} \\
%  $n_s$ & \num{0.9660499E+00 } \\
%  $r$ & \num{0} \\
%  \bottomrule
%\end{tabular}
%\end{center}}
%\caption{The physical parameters on which the CAMB calculated $\Lambda$CDM cosmology radiative transfer kernel is based on. They are used as presented in the Planck release \cite{Aghanim:2018eyx} from Table X, Column $Y$. }
%\label{table:phy_param}
%\end{table}

\begin{table}
\emph{\begin{center}
%s\noindent
\begin{tabular}{ l r r l }
  \toprule
  Parameter & \multicolumn{1}{c}{Planck Best-Fit} \\
  \midrule
  $H_0$ & \num{ 67.36} \\
  $\Omega_b h^2$ & \num{ 0.02237 } \\
  $\Omega_c h^2$ & \num{ 0.1200 } \\
  $m_\nu$ & \num{0.06 } eV \\
  $\Omega_K$ & \num{ 0 } \\
  $\tau$ & \num{ 0.0544 } \\
  $A_s$ & \num{ 0.9 } \\
  $n_s$ & \num{ 0.9649 } \\
  $r$ & \num{ 0 } \\
  \bottomrule
\end{tabular}
\end{center}}
\caption{The physical parameters on which the CAMB calculated $\Lambda$CDM cosmology radiative transfer kernel is based on. They are used as presented in the Planck release \cite{Aghanim:2018eyx} from Table X, Column $Y$. }
\label{table:phy_param}
\end{table}

\begin{figure}
%%\centering % \begin{center}/\end{center} takes some additional vertical space
\begin{subfigure}{1\textwidth}
\includegraphics[width=1.00\linewidth]{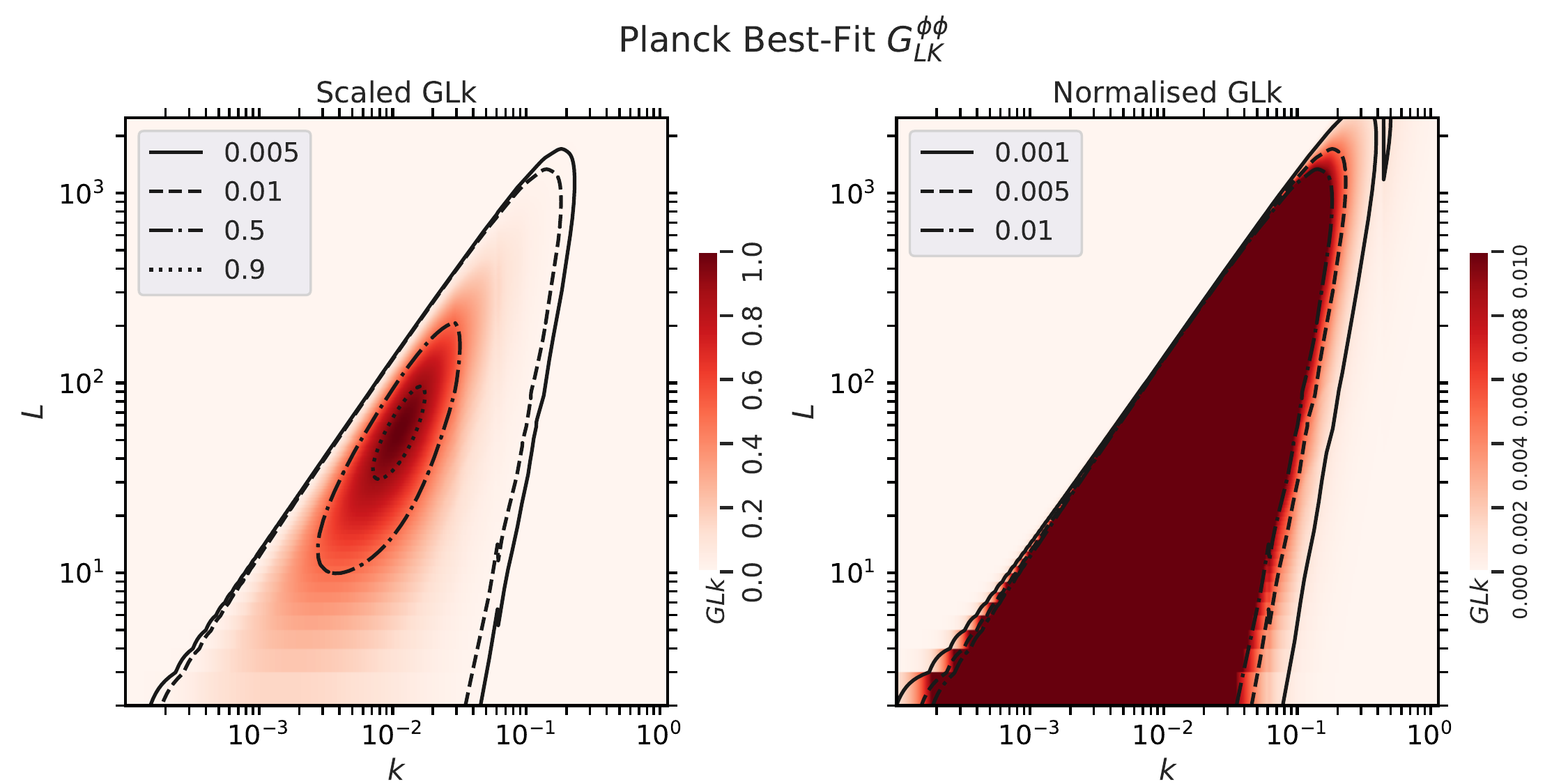}
%\caption{First subfigure} 
%\label{fig:kmaxcut_a}
\end{subfigure}
\caption{This plot shows a topdown heatmap of the scaled transfer kernel $G_L(k)/G_L(k)_{max}$. The first plot shows the kernel with the full range 0 to 1 on the heatbar and 4 power contours. The second plot shows the kernel heatmap zoomed into a range from 0 to 0.01 on the heatbar to show the extent of usable power that we can utilize to reconstruct $P_R(k)$, with corresponding contours. }
%{The plot shows the power that $G_L(k)$ transfers to $C_L^{\kappa\kappa}$ from different $P_R(k)$. It is a top down heatmap where the gradient bar in the legend varies from 0\% power to 0.001\% of the peak power of $G_L(k)$ (not normalized). The  acceptance criteria for the reconstruction $k_{max}$ is based on the Figure and analysis of \ref{fig:kmaxcut_b}, we put a cutoff of $k_{max}=10^0$ }
\label{fig:glk_heatmap_dual}
\end{figure}

%\begin{figure}
%%\centering % \begin{center}/\end{center} takes some additional vertical space
%\begin{subfigure}{1\textwidth}
%\includegraphics[width=1.00\linewidth]{Images/k_max_range.pdf}
%\caption{First subfigure} 
%\label{fig:kmaxcut_b}
%\end{subfigure}
%\hspace*{\fill} % separation between the subfigures
%\begin{subfigure}{1\textwidth}
%\includegraphics[width=1.00\linewidth]{Images/GLPP_3D.pdf}
%% "\includegraphics" is very powerful; the graphicx package is already loaded
%%\caption{Second subfigure} 
%\label{fig:kmaxcut_b}
%\end{subfigure}
%\caption{The two plots show how the amount of power that $P_{R}(k)$ contributes to $C_L^{\phi\phi}$ at different $k_{max}$ varies, parametrized by the expression  $k_{max} \propto 10^{\alpha}$ with the power as a parameter $\alpha$ }
%\label{fig:kmaxcut_b}
%\end{figure}

Since the Limber approximation is accurate mainly at high $L$, we divide the transport kernel calculation into two parts. First we calculate the kernel for an $L$ range $L \in [2, 100]$ where we use the exact expression in equation \eqref{eq:clppexact}. We first calculate the transfer function $T_{\phi}(k;\eta_0-\chi)$ from CAMB, then we integrate the function with the lensing window function and the spherical Bessel function of the first kind over $\chi$. This section takes approximately an hour to compute. For the $L$ range $L \in [101, 2500]$ we use the Limber approximated kernel equation \eqref{eq:clpplimber} to calculate the kernel and speed up the calculation over the intermediate steps where by interpolating over $\chi_{s}$ since it is now a function of $L$. The accuracy of the reconstruction can be verified by comparing the $C_L^{\phi\phi}$ constructed from kernel transport $C_{L}^{\kappa\kappa} =  \sum_{k} G_{Lk} P_R(k)$ to the $CAMB$ simulated $C_{L}^{\kappa\kappa}$ for the same physical parameters.  
%\textcolor{red}{Plot needed ???}

\begin{figure}[htb]
%%\centering % \begin{center}/\end{center} takes some additional vertical space
\begin{subfigure}{1\textwidth}
\includegraphics[width=1.00\linewidth]{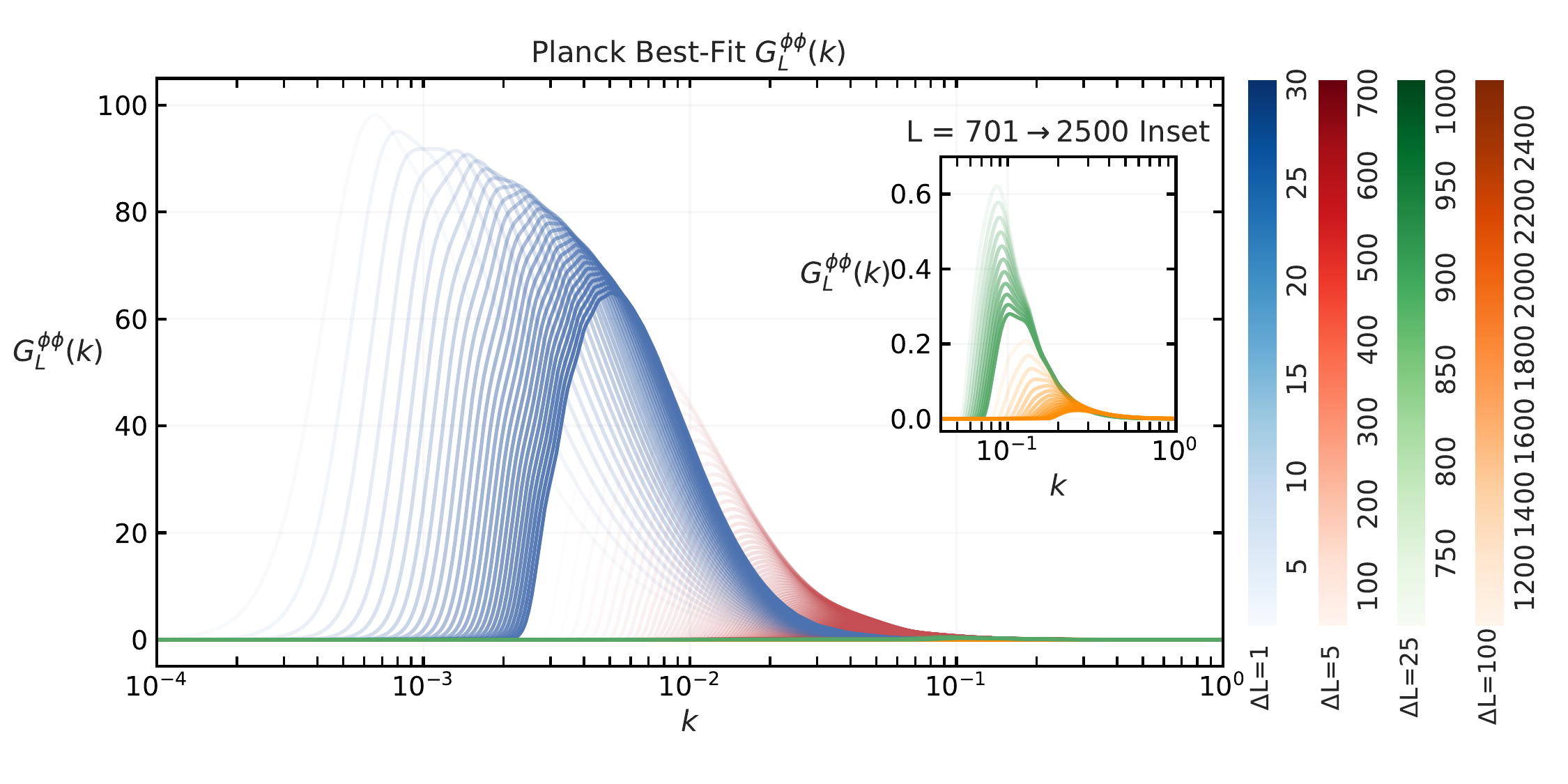}
\caption{} 
\label{fig:glpp_a}
\end{subfigure}
\hspace*{\fill} % separation between the subfigures
\begin{subfigure}{1\textwidth}
\includegraphics[width=1.00\linewidth]{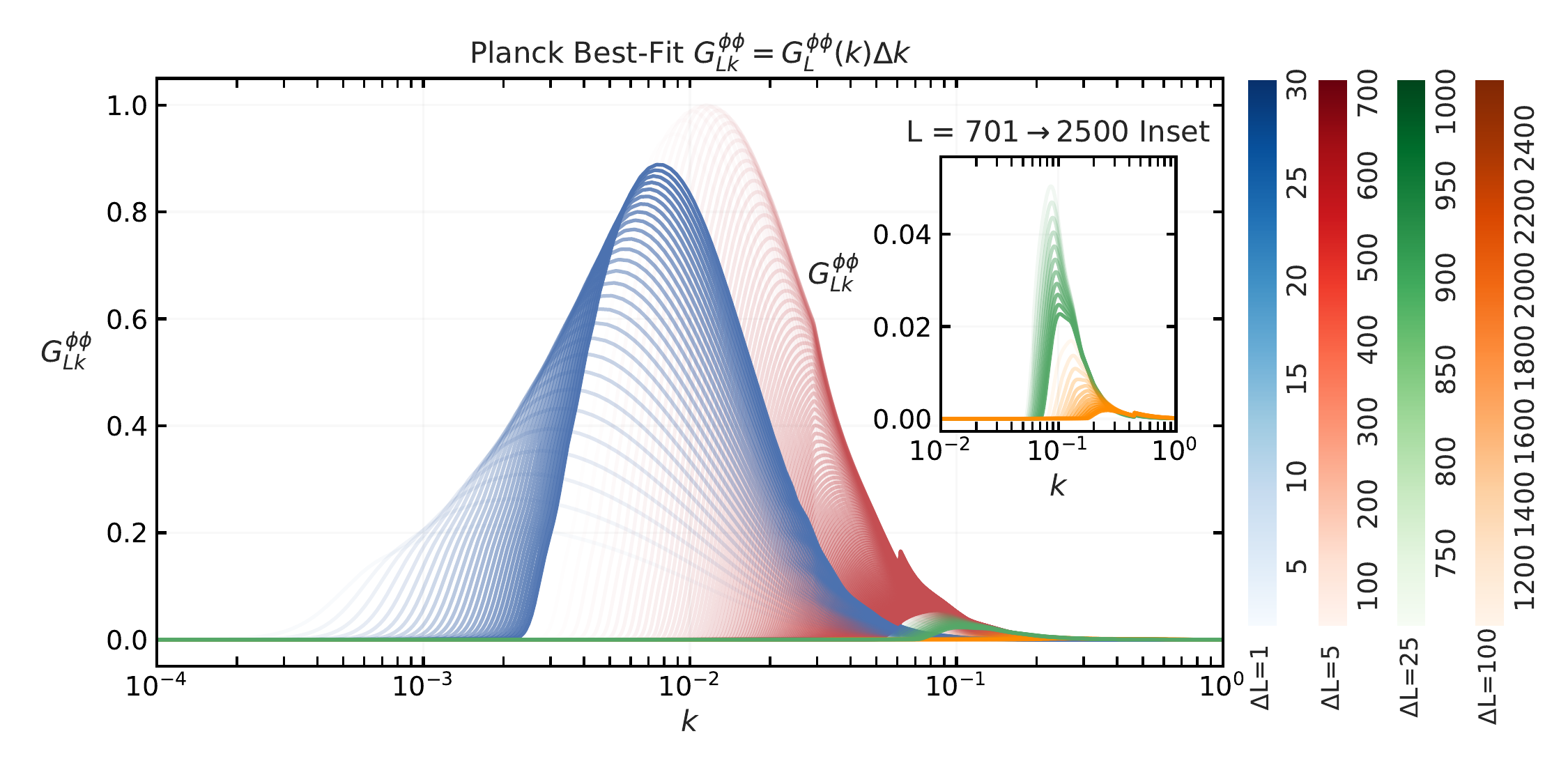}
%% "\includegraphics" is very powerful; the graphicx package is already loaded
\caption{} 
\label{fig:glpp_b}
\end{subfigure}
\caption{Plot \protect\subref{fig:glpp_a}) shows the functional form of the $G_{L}^{\phi\phi}{k}$ kernel projected onto the $K$ vs $G_{L}^{\phi\phi}(K)$ plane where $L$ is parametrized as a gradient in blocks of $L$ with corresponding $\Delta L$ step sizes informing the plotted frequency of $L$. The $L$ blocks are roughly segemented the amount of power they transfer. Plot \protect\subref{fig:glpp_b}) shows the same kernel but with the numerical integration step size multiplied, $G_{Lk}^{\phi\phi} = G_{L}^{\phi\phi}(k)\Delta{k}$. This helps visualise better how power at different $k$ is transferred to the angular power spectrum. }
\label{fig:glpp}
\end{figure}

\begin{figure}
\centering % \begin{center}/\end{center} takes some additional vertical space
\includegraphics[width=1.00\linewidth]{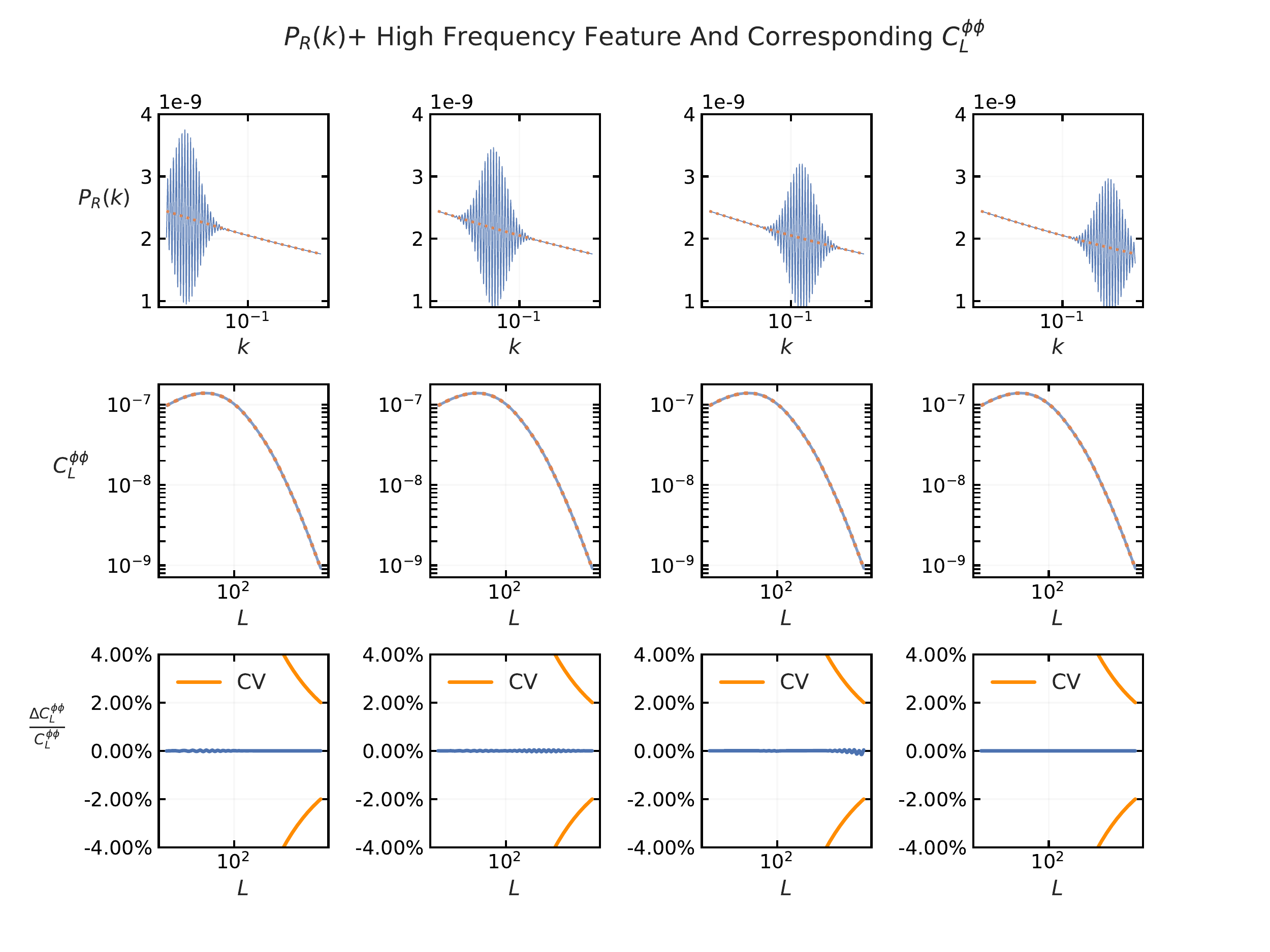}
\caption{This plot shows how a high frequency feature in the form of a wavepacket, plotted on the top row, has a net negligible contribution to $C_L^{\phi\phi}$, plotted in the mid row, with percent difference shown in the bottom row, which are all below cosmic variance. The $k$ ranges from $10^{-5}$ to $10$ and $L$ ranges from $2$ to $2500$}
\label{fig:pkhighfreq}
\end{figure}

The results of this simulation are expressed in figures \ref{fig:glk_heatmap_dual} and \ref{fig:glpp}. We can clearly see that the kernel is extremely smooth throughout the $k$ range over the entire range of $L$, especially at the range where most of the transfer power is present. This can lead us to conclude that any features in $P_{R}(K)$ that have a frequency higher than the radiative transfer kernel over the $k$ range will likely be degenerate with features that are smoother, yet transfer the same power to the resultant $C_L$. This means that reconstruction of very sharp features in $P_{R}(K)$ such as spikes or oscillatory features with a frequency higher than the kernel, will not be optimal or may be missed altogether. Test example of such features and their effect on $C_L^{\phi\phi}$ have been plotted in figure~\ref{fig:pkhighfreq}. As a result, we can also make some simplifications in the reconstruction process. We need not have a very fine grid over $k$ since the kernel is unable to reconstruct features with very high frequency in $k$. This reduces both our computation time as well as number of free parameters. For the sake of generality, in this paper we keep a fairly fine $k$ grid, higher than the expected reconstruction fidelity over $k$, but low enough for it to be less computationally heavy. In the future if we sample over the cosmological parameter space to find the best parameters for a free form $P_{R}(k)$, the freedom to reduce the grid density will aid in reducing the kernel computation time and making such an exercise computationally feasible. We provide an algorithm to optimize the $P_{R}(K)$ sampling in order to improve the statistical properties of the reconstructed solutions.
%\textcolor{red}{Figure for spike vs bump, leading to degenerate change in CL ???}

We observe the transfer function's properties along the $k$ ranges and their contribution across different $L$ multipoles. Figure \ref{fig:glk_heatmap_dual} shows a top down heatmap of the transfer kernel $G_{Lk}^{\phi\phi}$

\begin{figure}
\centering % \begin{center}/\end{center} takes some additional vertical space
\includegraphics[width=1.00\linewidth]{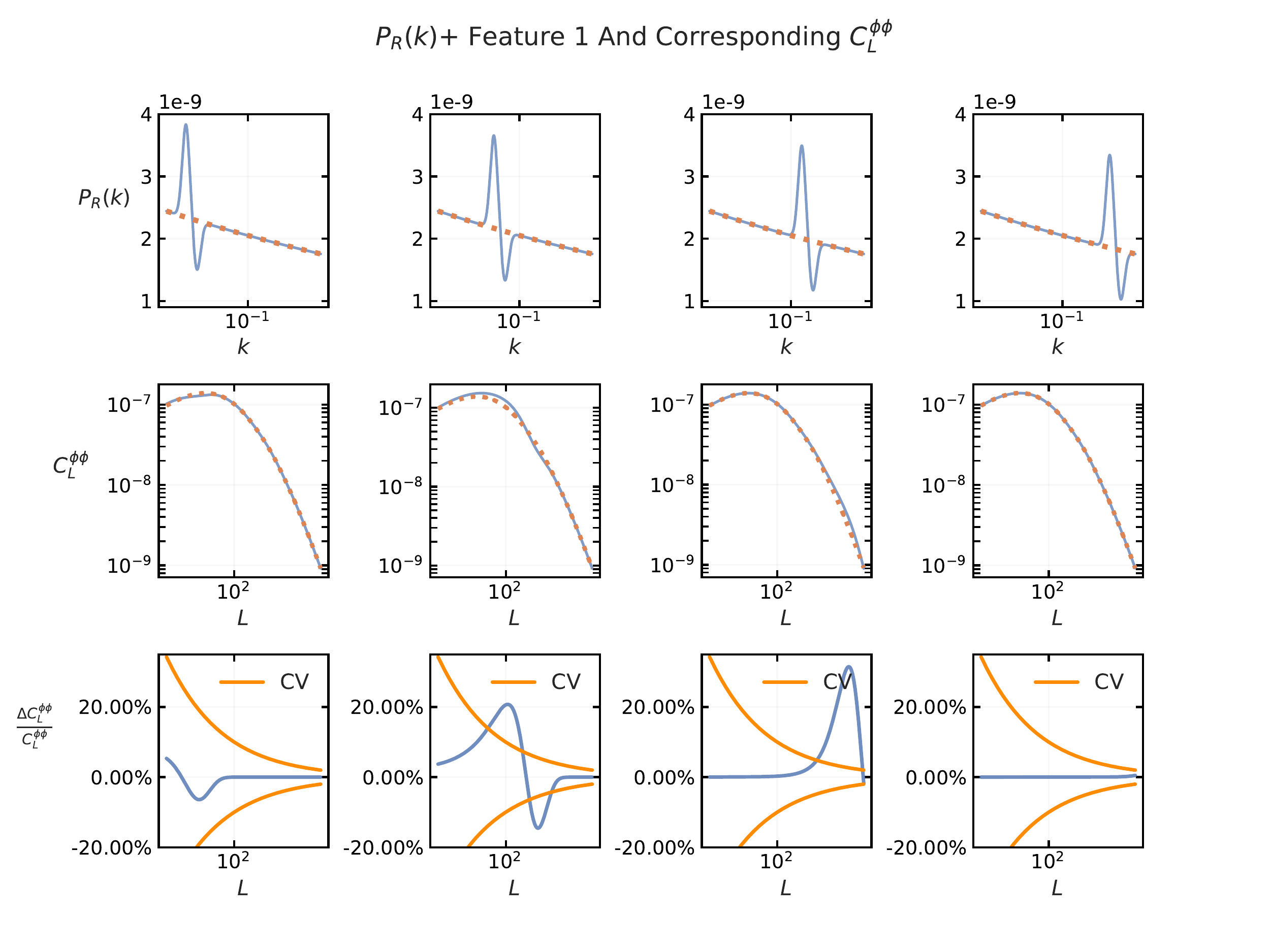}
\caption{This plot shows how a low frequency feature, Feature 1, plotted on the top row, has a varying contribution to $C_L^{\phi\phi}$ as it moves across the $k$ range, with the corresponding $C_L^{\phi\phi}$ plotted in the mid row, with percent power difference shown in the bottom row, which vary according to the kernel~\ref{fig:glpp}. The $k$ ranges from $5\times10^{-4}$ to $10$ and $L$ ranges from $8$ to $2500$}
\label{fig:pkhFeature1}
\end{figure}

We can see that most of the kernel transfer power lies in the range of $k \in 10^{-3} \rightarrow 10^{-1} $. This means that the reconstruction support of the Richardson-Lucy estimator lies primarily in this range and smooth features in $P_{R}(k)$ within this region are more likely to be reconstructed with higher fidelity than in the regions beyond this range. Depending on the minimum $L$ range of the observed data, this will typically lead to the previously mentioned range of $k \in 10^{-3} \rightarrow 10 $. Hence in short, this is the region where any $P_{R}(k)$ estimation from $C_L^{\phi\phi}$ is likely to be meaningful. To verify this assumption, we plot the $C_L^{\phi\phi}$ constructed by a $P_{R}(k)$ superimposed with some relatively low frequency feature, which we shall term Feature 1 and study how its contribution varies across the $k$ location of the feature. In Figure~\ref{fig:pkhFeature1} a simulated low frequency superimposed feature which we call Feature 1, is plotted moving across the $k$ range over $P_{R}(k)$ and its contribution convolved with the transfer function is obtained as  $C_L^{\phi\phi}$ and the power differential with respect to a vanilla power law $P_{R}(k)$. As surmised earlier, the main contribution comes primarily when the feature moves across the $10^{-2}$ mark which coincides with the kernel peak in Figure~\ref{fig:glpp}.

% PK Feature 1 travel across k range 

Now we explain the numerical properties of the kernel and the role they play in optimal reconstruction of $P_R(k)$.
As we can see from Figure~\ref{fig:glpp_b}, to make reconstruction efficient and optimal, we primarily need to assign an equal number of $P_R(k)$ samples with respect to the number of $C_L^{\phi\phi}$ data points. Fundamentally we are solving a set of linear equations and in order to have a well behaved covariance matrix, we need the system to be exactly determined and consistent set of linear equations. We can also work with an underdetermined and consistent set of equations where the $k$ sampling is much higher than the number $L$ samples, but this will lead to issues, which will be elaborated in the reconstruction sections.
Apart from this consideration, we also need to appreciate that the behaviour of the transport kernel $G_{LK}^{\phi\phi}$ is such that a given $k$ samples can be reconstructed better if it contributes to different $L$ at different strengths. If it contributes to several $L$ points at the same strength, this would imply a degeneracy and make it difficult to distinguish the $P_R(k)$ at that $k$ from a different $k$ doing the same. Recall that the core of the R-L algorithm is the $G_{LK}^{\phi\phi}$ kernel normalised along $L$, effectively weighing the reconstructed $P_R(k)$ with respect to their contribution to different $L$. 
Based on these discussions, we propose a mechanism where we establish a minimum power cutoff, $G_{LK cutoff}^{\phi\phi} = 0.001\% G_{LK max}^{\phi\phi}$.
Then we can divide the kernel into several $k$ sections and selectively choose $k$ samples out of the simulation superset, based on how many $L$ modes 'enter' the $G_{LK cutoff}^{\phi\phi}$ threshold within each section. This will take care of both the constraint of having $N_k = N_L$ as well as ensuring that a given $k$ samples is tagged with a corresponding and unique $L$ mode to ensure that more 'staggered' regions of the kernel are utilized for reconstruction, instead of the homogenous regions, where it will be difficult to distinguish one $k$ contribution from a neighbouring one.
In addition, based on the $G_{LK cutoff}^{\phi\phi}$ and with reference to the contour plot \ref{fig:glk_heatmap_dual} , we will also cut off any $k$ and $L$ regions that fall outside the cutoff contour. We provide a general idea of the reconstruction regions in the table \ref{table:k_section}.

\begin{table}
\emph{\begin{center}
%s\noindent
\begin{tabular}{ l r r l }
  \toprule
  $k$ range & \multicolumn{1}{c}{$L$ range} \\
  \midrule
  $k \in [2 \times 10^{-4} \rightarrow 2 \times 10^{-3}] $ & $L \in [2 \rightarrow 30]$ \\
  $k \in [2 \times 10^{-3} \rightarrow 5\times 10^{-2}] $ & $L \in [31 \rightarrow 700] $ \\
  $k \in [5\times 10^{-2} \rightarrow 2\times 10^{-1}] $ & $L \in [701 \rightarrow 1350] $\\
  $k \in [2\times 10^{-1} \rightarrow 1\times 10^{0}] $ & $L \in [1350 \rightarrow 2500] $\\
  \bottomrule
\end{tabular}
\end{center}}
\caption{The reconstruction regimes ordered by Data $C_L$ vs Reconstruction $P_R(k)$ degrees of freedom $L$ and $k$.}
\label{table:k_section}
\end{table}

Another observation to note is that most of the kernel power lies within an $L$ range of $L \in 2 \rightarrow 700 $, This means that any features in $P_{R}(K)$ will reflect in the $C_L^{\phi\phi}$ more significantly within this $L$ region. As a result, precise data in this $L$ range is likely to be more valuable in context of $P_{R}(K)$ deconvolution. Since a significant portion of the kernel support transfers power to the low $L$s, this calls for high precision full sky reconstructions of $C_L^{\phi\phi}$. To this end we shall examine how Cosmic Variance limited data would affect reconstruction and the discriminatory precision in extracting features in $P_{R}(K)$ or lack thereof. 
%Figure~\ref{fig:Feature_ReconRlzn_Unbin_Bin} shows the reconstruction obtained using cosmic variance limited simulated data. 
%\textcolor{red}{Figure for CV limited unbinned and binned CLPP, its reconstruction with $\sigma$ bands}

\subsection{Initial Guess Behaviour}
\label{sub_sec:init_guess_behav}

In this section we further expand upon the effects of the kernel on IRL reconstruction and the initial guess $P_{R}(K)^{(i=0)}$ used to begin the IRL iterations. Using the CAMB generated Limber approximated $C_L^{\kappa\kappa}$ data as before, we reconstruct the input Power-Law $P_{R}(K)$ with different initial guesses $P_{R}(K)^{(i=0)}$ and observe how the reconstruction varies. Ideally we should expect no change as the injected data does not change in either iteration.

\begin{figure}[htb]
%%\centering % \begin{center}/\end{center} takes some additional vertical space
\begin{subfigure}{1\textwidth}
\includegraphics[width=1.00\linewidth]{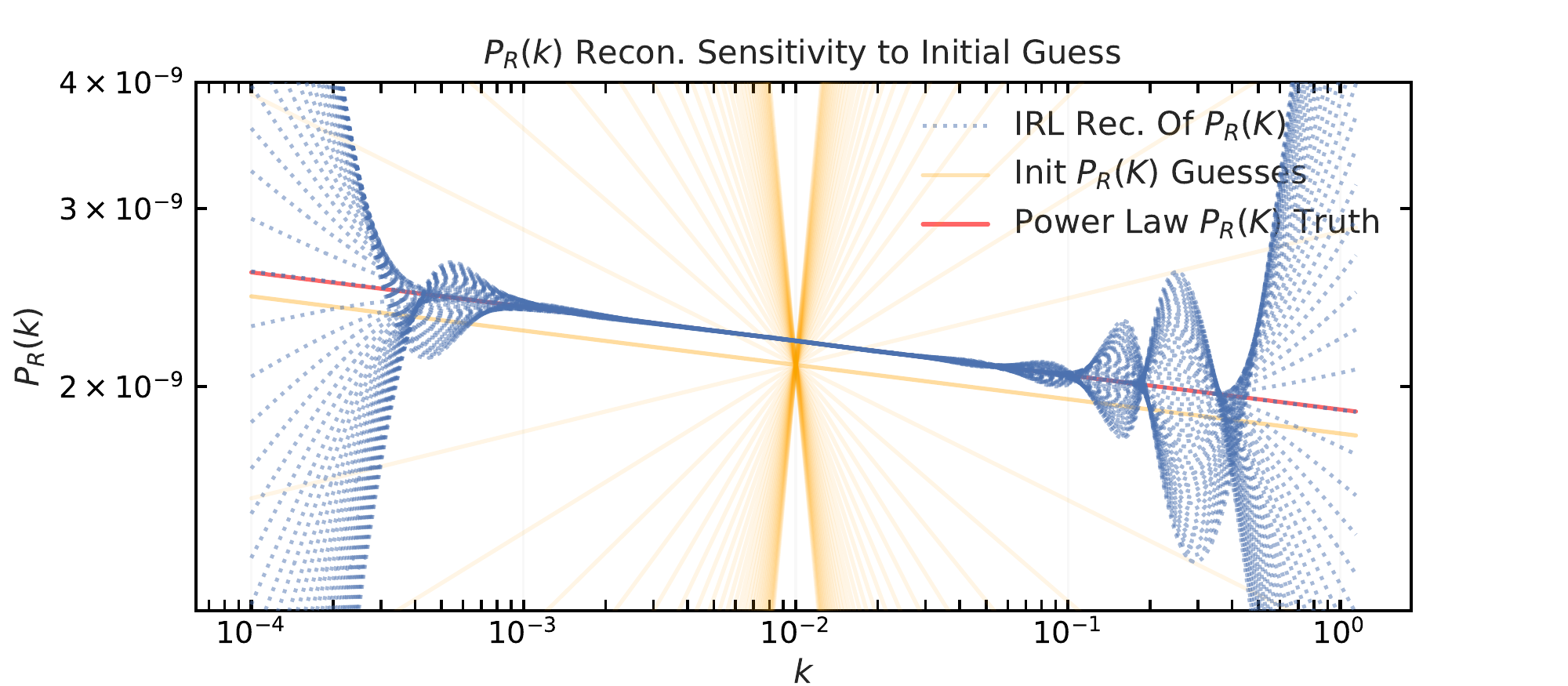}
%% "\includegraphics" is very powerful; the graphicx package is already loaded
\caption{} 
\label{fig:slope_1}
\end{subfigure}
\hspace*{\fill} % separation between the subfigures
\begin{subfigure}{1\textwidth}
\includegraphics[width=1.00\linewidth]{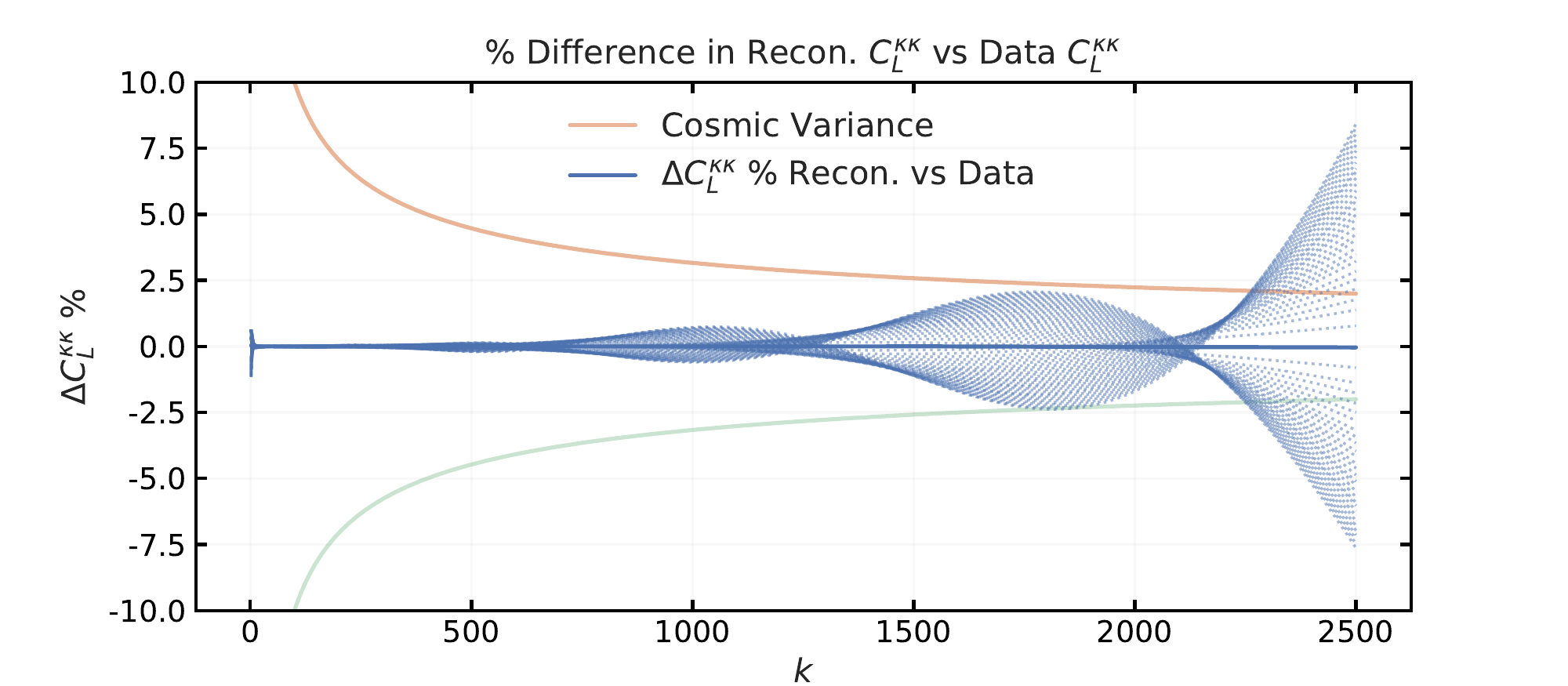}
%% "\includegraphics" is very powerful; the graphicx package is already loaded
\caption{} 
\label{fig:slope_2}
\end{subfigure}
\caption{Figure \protect\subref{fig:slope_1}) shows the reconstructed $P_{R}(k)$ in blue dashed lines, given different initial guesses $P_{R}(k)^{(i=0)}$ in yellow lines varying by slope $n_s$. The red lines shows the original injected power spectrum $P_{R}(K)$. Figure \protect\subref{fig:slope_2}) shows the relative \% difference in the reconstructed $\hat{C}_L^{\kappa\kappa}$ and the input data $C_L^{\kappa\kappa}$. The oscillations due to different reconstructions show how the kernel is largely insensitive at high $k$ and hence high $L$ and variations are dominated by the initial guess. }
\label{fig:init_guess_slope}
\end{figure}

Figure \ref{fig:init_guess_slope} shows how a range of different initial guesses $P_{R}(K)^{(i=0)}$ obtained by varying the slope of the initial guess $P_{R}(K)^{(i=0)} = A_s(\frac{k}{k*})^{(n_s-\frac{j}{10})}$ for integer $j\in[-20\rightarrow40]$ recover the original $P_{R}(K)$. It is clear from the plot that slope of the initial guess has a large impact on the reconstruction at high and low $k$'s. For low $k$ this does not have much effect in the final reconstruction as the error bars in this region will too large anyway to affect feature hunting. But at high $k$, we impose a reconstruction bound of $k_{max} = 0.2$ when feature hunting and calculating the $\chi^{2}$ of the reconstructed $P_{R}(K)$ to distinguish it from the null hypothesis power law. Usually we can use an initial guess who slope is as drastic as some of the examples used in the Figure, this in turn allows for a control over the reconstruction at edge $k$'s. But this is equivalent to imposing some prior information, in that we expect the overall slope of any reconstructed $P_{R}(K)$ to be close to the fiducial $n_s$ and only hunt for localized features 'riding' a global power law. Relaxing such limitations can be looked at in detail in future work. We also provide some more initial guess variation plots in figures \ref{fig:init_guess_up} and \ref{fig:init_guess_down} to show how the reconstruction is highly dependent on the initial guess slope at high and low $k$'s.

This section essentially informs us of the limitations of the kernel and the IRL estimator in the low support regions of the transfer kernel and hence which $k$ ranges we can reliably use for feature hunting and $P_{R}(K)$ reconstruction.

% CV Limited Feature 1 : UpDown realisation using unbinned and Planck bin size binned data.
\begin{figure}[htb]
    \centering % <-- added
\begin{subfigure}{0.5\textwidth}
  \includegraphics[width=\linewidth]{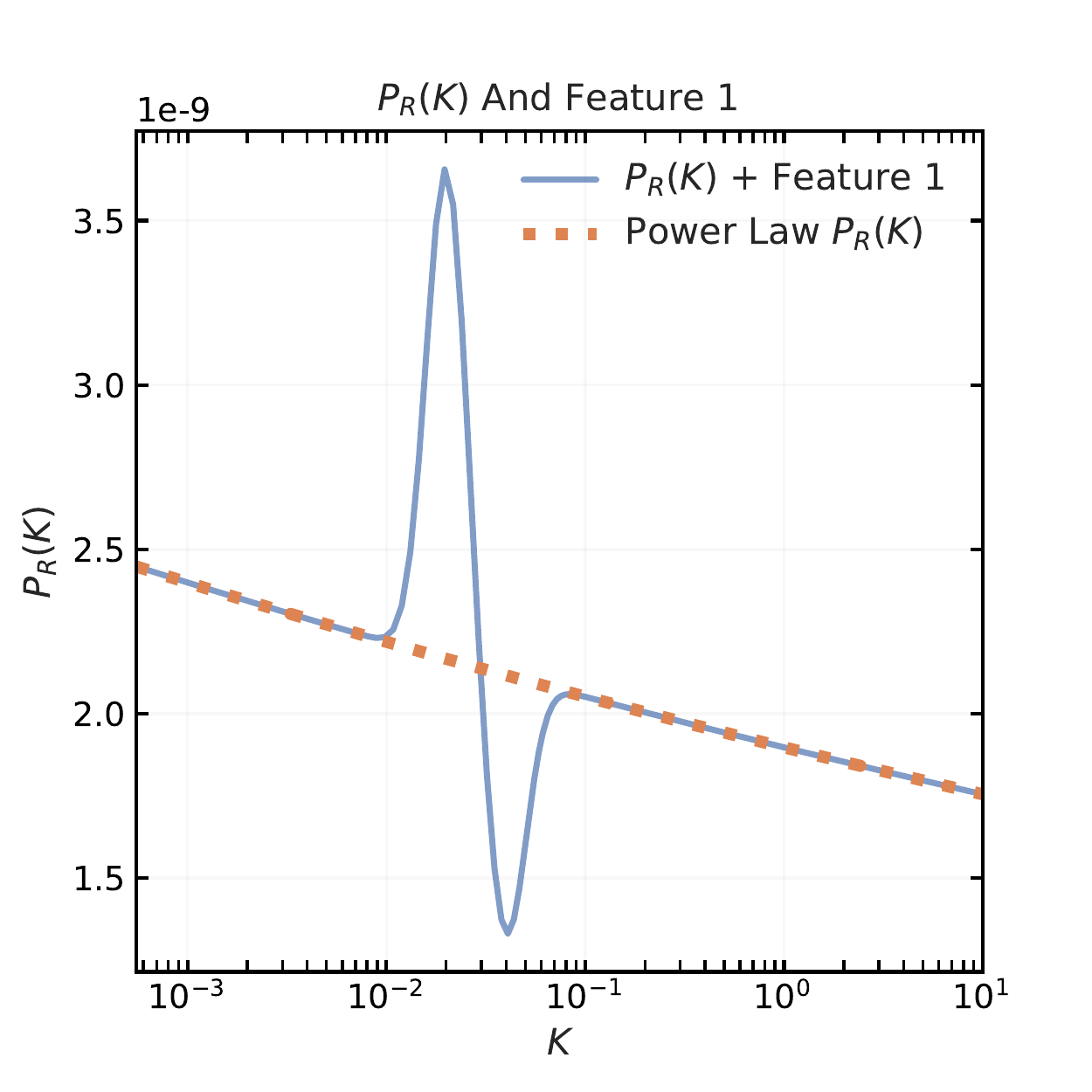}
  \caption{}
  \label{fig:Feature_DataRlzn1}
\end{subfigure}\hfil % <-- added
\begin{subfigure}{0.5\textwidth}
  \includegraphics[width=\linewidth]{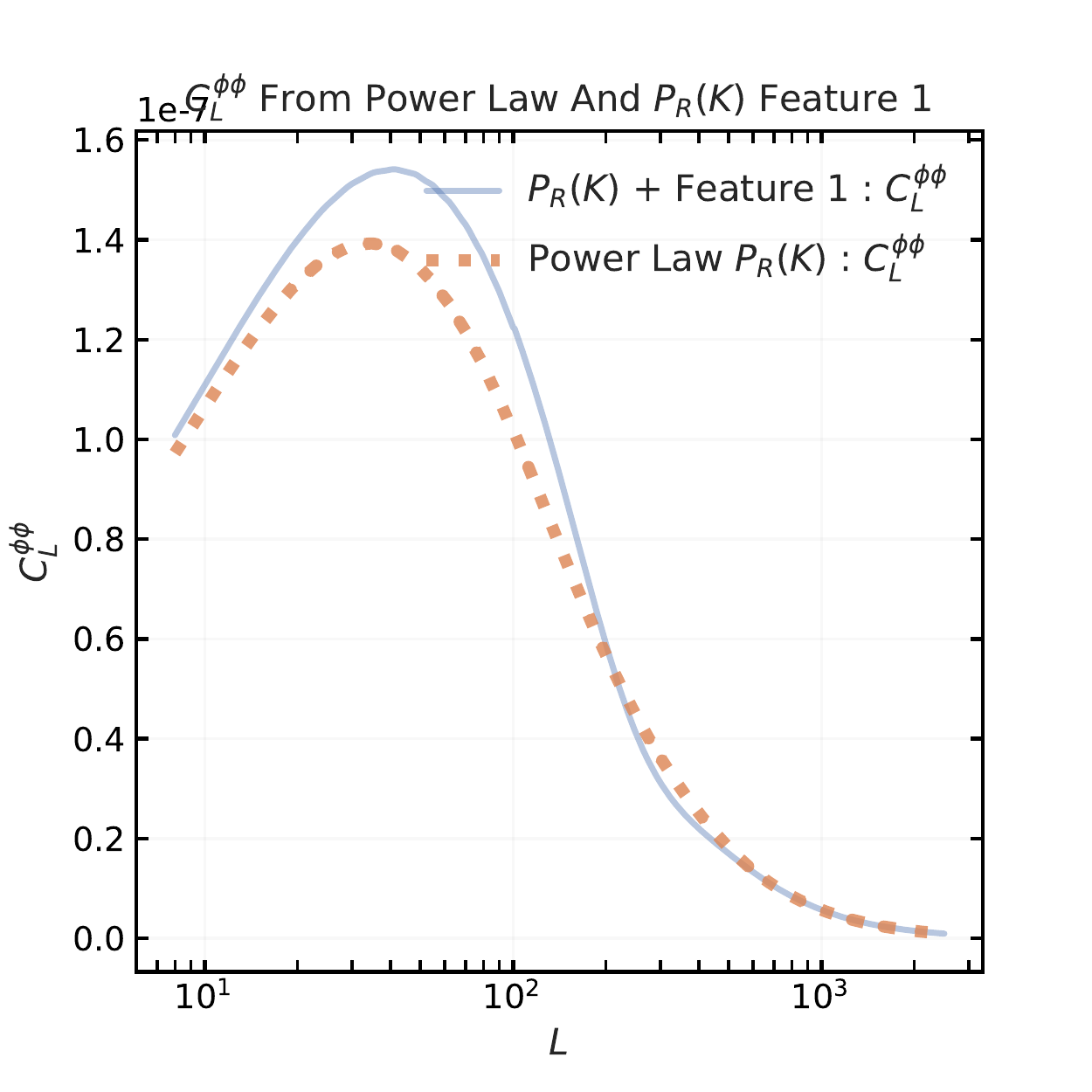}
  \caption{}
  \label{fig:Feature_DataRlzn2}
\end{subfigure}
\caption{\protect\subref{fig:Feature_DataRlzn1}) Plots a low frequency feature, Feature 1, superimposed on the power law $P_{R}(k)$. \protect\subref{fig:Feature_DataRlzn2}) Plots the corresponding $C_{L}^{\phi\phi}$ from $P_{R}(k)$ with and without Feature 1.}
\label{fig:Feature_DataRlzn}
\end{figure}

% CV Limited Feature 1 : UpDown reconstruction using unbinned and Planck bin size binned data.

\section{Results Ib : Simulated IRL Reconstruction}
\label{sec:ressimtest}

%1. Should I do all reconstruction with CV or with artifical error bars based on experiment projections ?	\\
%2. One set of reconstruction on pure power law PRk based on best fit parameters used to make kernel as well. See the ideal recon properties and cov mat, chisqstat.	\\
%3. One set of reconstruction on a PRk with some feature ?	\\
%4. 1,000,000 nrlzn for cov mat pos def since nrlzn > nvar**2	\\
%5. Use the convergence criteria decided to stop recon.	\\
%6. Find the minimum strength of a feature that can be detected at (2/3+)$\sigma$ level in PRk reconstruction in the highest precision region of $k$. This determines optimal performance of estimator given data and error bars.	\\
%7. \\

In this section we perform a reconstruction of the PPS on simulated data with appropriate error bars, on the unbinned data, as well as analyze the error budget and covariance properties of the reconstruction. This seeks to provide a good foundation to implement the process on observed Planck data and predict the science output of future observations analyzed by this reconstruction process.

\subsection{Results i) : High $k$ Density}
\label{sub_sec:fullksim}

For the primary reconstruction example, we demonstrate the application of the IRL reconstruction algorithm on the the base $\Lambda$CDM cosmology model based on Planck best fit parameters and the Inflationary Primordial Power Spectrum $P_{R}(k)$ using a power law model $A_s(k/k_*)^{(n_s-1)}$. We list the simulation parameters and conditions in the following table \ref{table:unbin_IRL_powlaw}.

\begin{table}
%\emph
{\begin{center}
%s\noindent
\begin{tabular}{ l r r l }
  \toprule
  Parameter & \multicolumn{1}{c}{Values} \\
  \midrule
  Input Data & $C_L^{\kappa\kappa}$ \\
  Error Bars & Cosmic Variance \\
  Input $P_{R}(k)$ Model & $A_s(k/k_*)^{(n_s-1)}$ \\
  $k$ Range & [ \num{e-4} $\rightarrow$ \num{1.14} ] \\
  $N_k$ & 1853 \\
  $L$ Range & [ \num{2} $\rightarrow$ \num{2500} ]  \\
  $L$ Binning & Unbinned \\
  Data Realisations & \num{2e6} \\
  \bottomrule
\end{tabular}
\end{center}}
\caption{The simulation parameters over which the IRL reconstruction is carried out.}
\label{table:unbin_IRL_powlaw}
\end{table}

\begin{figure}
\centering % \begin{center}/\end{center} takes some additional vertical space
\includegraphics[width=1.00\linewidth]{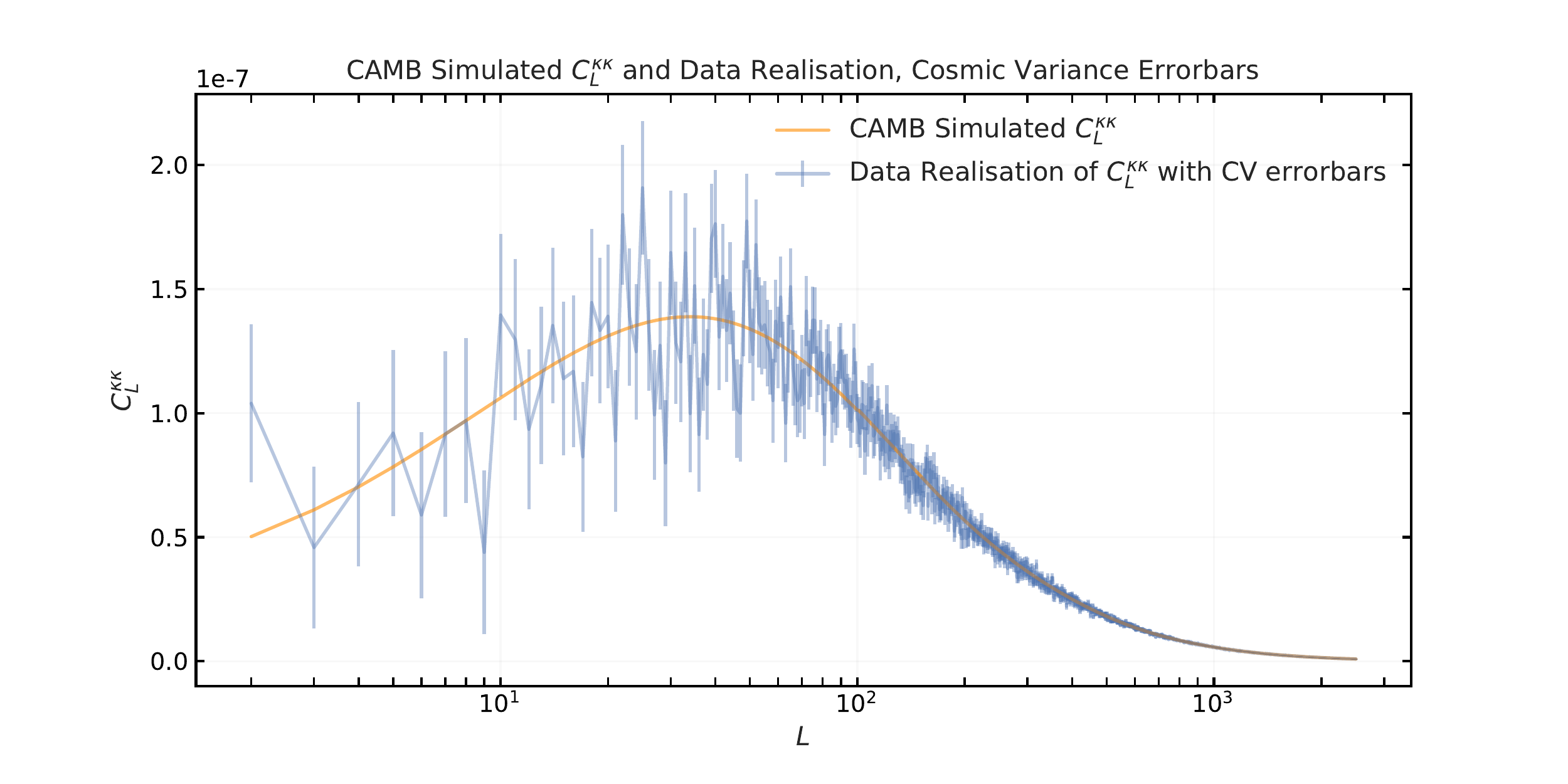}
\caption{Plot of the CAMB simulated $C_L^{\kappa\kappa}$ in Orange and the data realisation by treating each $C_L^{\kappa\kappa}$ as a Gaussian random sample based on cosmic variance error bars, in blue.}
\label{fig:input_clkk_data}
\end{figure}

The input data is limited by cosmic variance limited errorbars, to study the ideal case for reconstruction demonstration. The CAMB simulated $C_L^{\kappa\kappa}$ power spectrum and its data realisation, are presented in figure \ref{fig:input_clkk_data}.

Since we are working with an order of magnitude $10^3$ number of free form $P_{R}(k)$ variables, we expect a sample size of $10^6$ realisations of these measured variables to provide a statistically meaningful reconstruction of the IRL reconstructed $P_{R}(k)$ error covariance matrix $\Sigma_{kk'}$. Based on the input data realisation and errorbars in figure \ref{fig:input_clkk_data}, we generate Monte-Carlo samples and carry out IRL reconstructions on them. We note that we are working with the error bars in $C_L^{\kappa\kappa}$ assuming that they are not correlated, meaning that $\Sigma_{LL'}$ is diagonal. Hence we will use the $2^{nd}$ form of the IRL algorithm in equation \ref{eq:IRL_eqns}. For real data usually this is not the case and the full $\Sigma_{LL'}$ may need to be incorporated. 
We provide a plot of the accuracy saturation of $\Sigma_{kk'}$ based on the IRL reconstructed $P_{R}(k)$ samples, to find the optimum realisation numbers required for accuracy. In the figure \ref{fig:Feature_DataRlzn} we plot the accuracy saturation of the reconstructed $\Sigma_{kk'}$ with respect to increasing realisation count and establish \num{2e6} as an acceptable number of realisations.

\begin{figure}[htb]
    \centering % <-- added
\begin{subfigure}{0.5\textwidth}
  \includegraphics[width=\linewidth]{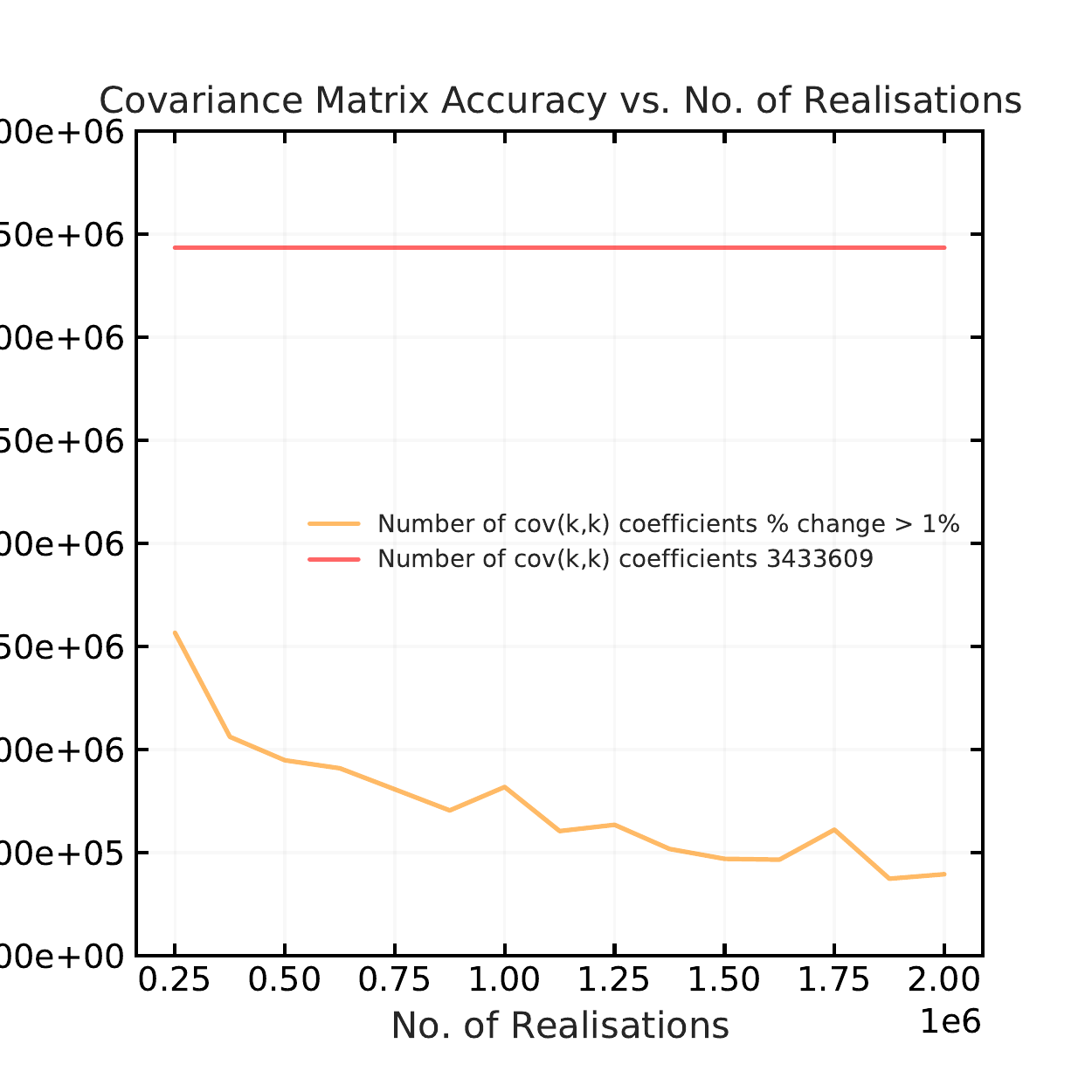}
  \caption{}
  \label{fig:1perc_sat}
\end{subfigure}\hfil % <-- added
\begin{subfigure}{0.5\textwidth}
  \includegraphics[width=\linewidth]{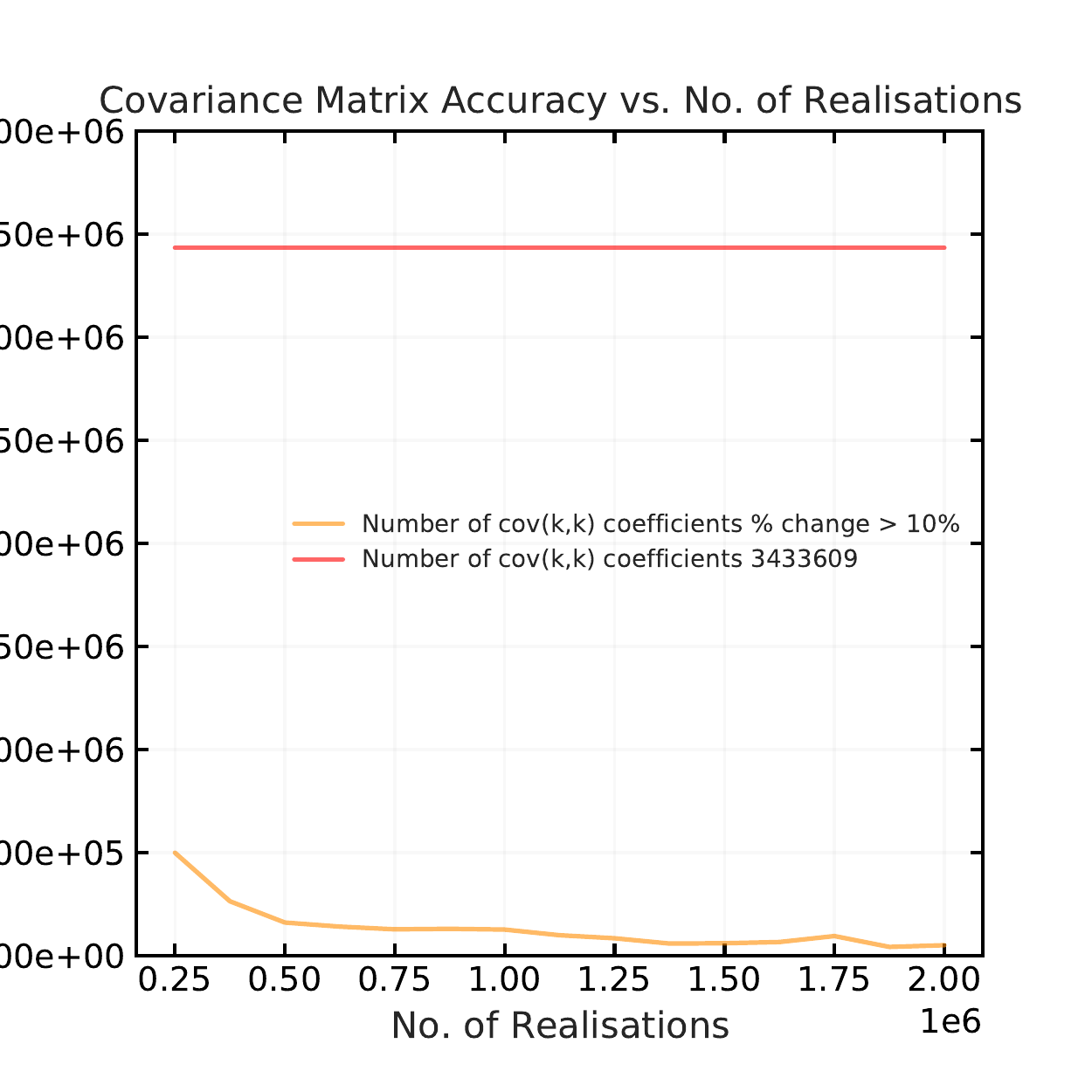}
  \caption{}
  \label{fig:10perc_sat}
\end{subfigure}
\caption{ \protect\subref{fig:1perc_sat}) Plots the number of $\Sigma_{kk'}$ coefficients that exceed a $1\%$ relative error change with respect to the previous realisation set, vs the realisation number set. We see that by \num{2e6} number of realisations, the accuracy saturates. Similarly the plot \protect\subref{fig:1perc_sat}) plots the same but at $10\%$ accuracy threshold, showing similar saturation.}
\label{fig:Feature_DataRlzn}
\end{figure}

\begin{figure}
\centering % \begin{center}/\end{center} takes some additional vertical space
\includegraphics[width=1.00\linewidth]{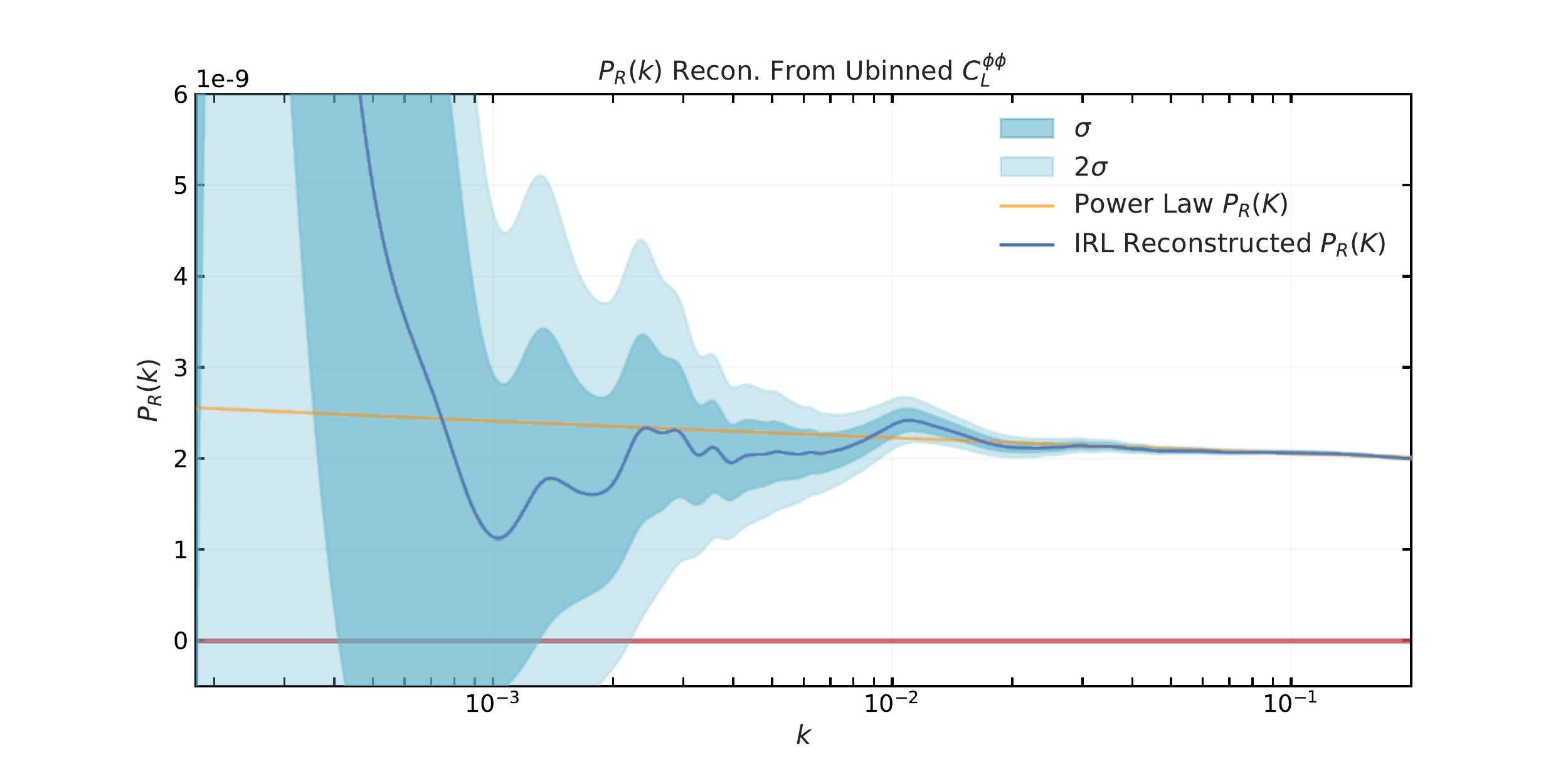}
\caption{Plot of the reconstructed $P_{R}(k)$ (Blue) with $1\sigma,2\sigma$ bands from $2\times10^{6}$ data realisations with cosmic variance error, overplotted on the fiducial Power Law $P_{R}(k)$ (Orange).}
\label{fig:recon_unbin_allk_pk_sigband}
\end{figure}

Based on these reconstructed realisations we plot the reconstructed $P_{R}(k)$ and the $1\sigma, 2\sigma$ error bands from the diagonal part of $\Sigma_{kk'}$, in plot \ref{fig:recon_unbin_allk_pk_sigband}.

The reconstruction clearly has varying accuracy with respect to $k$. This follows from the nature of the input $C_L^{\kappa\kappa}$ data. At low $L$, which correspond to power acquired from low $k$, (as can be noted from the kernel plots), the cosmic variance is high due to less number of azimuthal multipoles $N_m = 2L+1$. This means that the data realisation $C_L^{\kappa\kappa}$ has a higher degree of fluctuations that deviate from the fiducial CAMB $C_L^{\kappa\kappa}$, which are then transferred to the reconstructed $P_{R}(k)$ as fluctuations at low $k$. This behaviour makes it difficult to ascertain features in $P_{R}(k)$ at low $k$ due to the inherently noisy nature of the data. This can be addressed by aggressive binning at low $L$, which can clean the fluctuations, but has the downside of reducing the $k$ sampling at those $L$s. The reason for this is addressed in the $\Sigma_{kk'}$ matrix later; an inordinately high $k$ sampling can lead to degenerate solutions for $P_{R}(k)$ and high degree of correlation in the free form $P_{R}(k)$ reconstruction, again making recovery of features difficult as they may not be unique or independent of correlated features. Overall this leads to the conclusion that from low $L 
\in [2\rightarrow30]$, $k \in [10^{-4} \rightarrow 3\times10^{-3}]$, accurate, statistically uncorrelated sampling of $P_{R}(k)$ is sparse and not much information can be extracted over that range. What can be, gives a limited picture in terms of feature hunting.
The reconstruction improves significantly at mid range $L 
\in [31\rightarrow1200]$ corresponding to $k \in [ 3\times10^{-3} \rightarrow 10^{-1} ]$. See figure \ref{fig:glpp}. In this range a major chunk of the $C_L^{\kappa\kappa}$ data gets contribution from the said $k$ range and hence $P_{R}(k)$ can be sampled densely and will be largely uncorrelated, hence features detected in this range are relatively more statistically significant and unique. In addition the cosmic variance error bounds are significantly low here, so both the fluctuations in the data realisation are lesser, as well as propagated error bands in $P_{R}(k)$ are narrower, which means a higher precision reconstruction in this region make it easier to narrow down potentially interesting deviations from the Power Law model.
We propose that the estimator works well as a high precision reconstruction technique in this region of $k$ and is limited mainly by experimental bounds.
At higher $k \in [ 10^{-1}  \rightarrow 1]$ ranges, the reconstruction, though having a very low error bound due to low cosmic variance, is unreliable due to a high degree of correlation between the $P_{R}(k)$ samples. The discussion about the kernel in section \ref{sec:kernelsim} explains that above $k = 1.5\times10^{-1}$, the kernel becomes uniform enough over $L$ for the IRL algorithm to be unable to distinctly reconstruct individual $P_{R}(k)$ points. Furthermore the kernel support itself drops rapidly and a full $k$ range reconstruction here is numerically unsound, as shown in section \ref{sub_sec:init_guess_behav}. It may be possible to study what happens when limited high $k$ ranges are reconstructed independently by analysing specific $C_L^{\kappa\kappa}$ bands over $L$ and high $L$, but that work is reserved for future updates.

\begin{figure}
\centering % \begin{center}/\end{center} takes some additional vertical space
\includegraphics[width=1.00\linewidth]{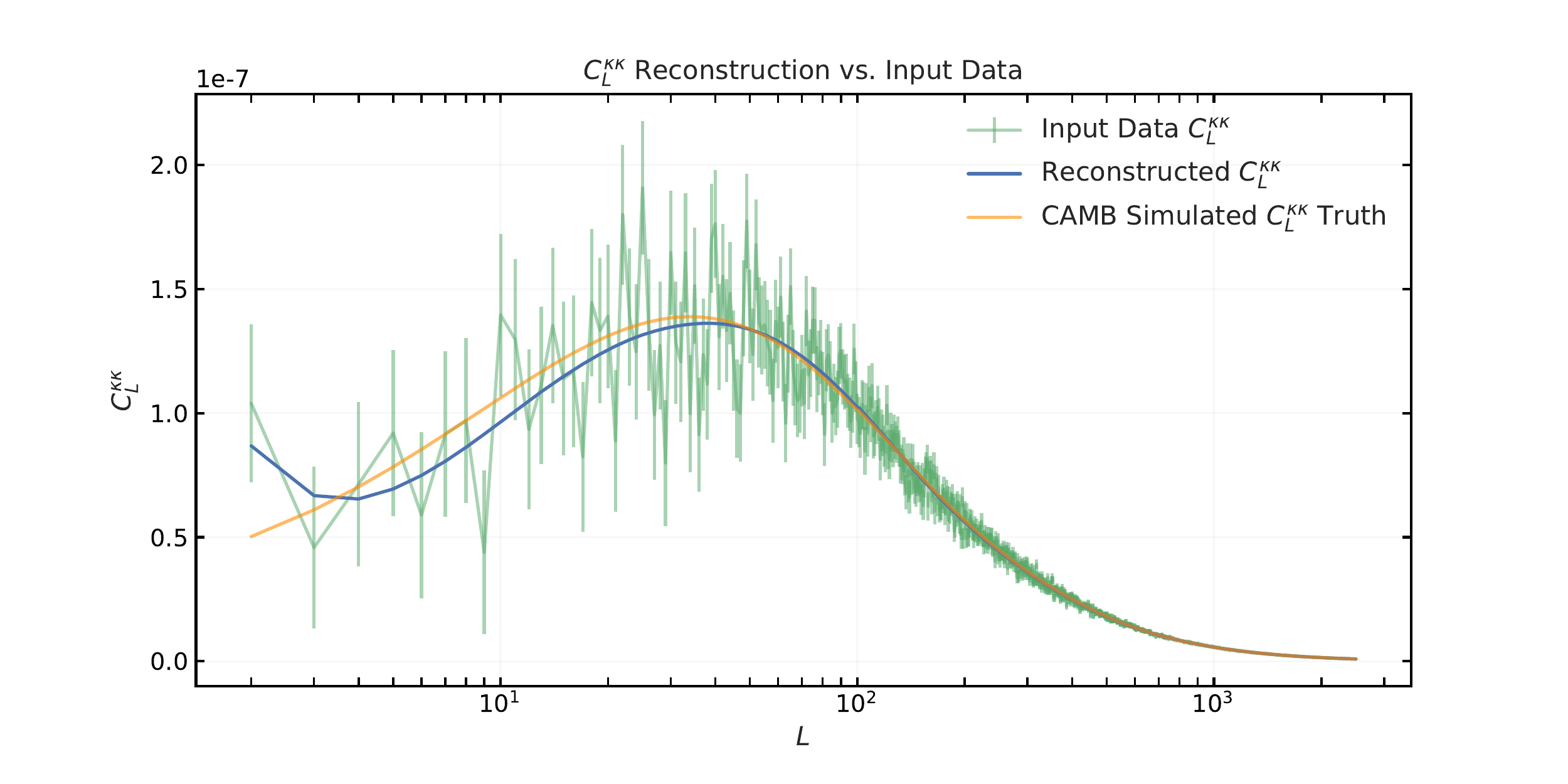}
\caption{Reconstructed $C_L^{\kappa\kappa}$ in Blue plotted over the fiducial CAMB $C_L^{\kappa\kappa}$ in Orange and the input data realisation in Green.}
\label{fig:recon_unbin_allk_clkk_vs_inp}
\end{figure}

\begin{figure}
\centering % \begin{center}/\end{center} takes some additional vertical space
\includegraphics[width=1.00\linewidth]{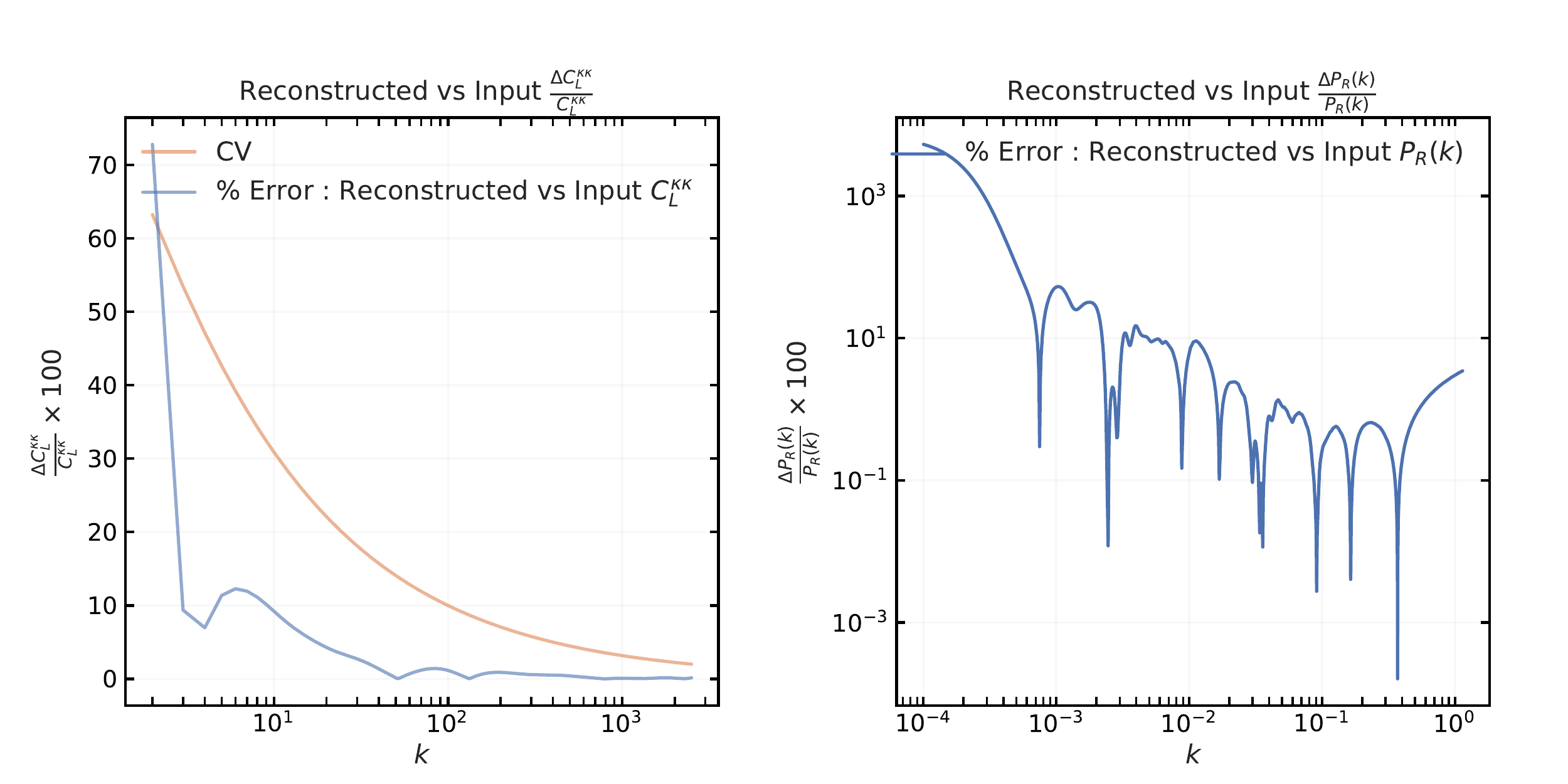}
\caption{The two figures show the relative \% error between the reconstructed $C_L^{\kappa\kappa}$ vs data realisation $C_L^{\kappa\kappa}$, and the reconstructed $P_{R}(k)$ vs input Power Law $P_{R}(k)$}
\label{fig:clkk_pk_rel_error}
\end{figure}

We also plot the reconstructed $C_L^{\kappa\kappa}$ and compare it to the input data and fiducial CAMB simulated $C_L^{\kappa\kappa}$ in figure \ref{fig:recon_unbin_allk_clkk_vs_inp}, along with the relative error plots between both reconstructed $C_L^{\kappa\kappa}$ and $P_{R}(k)$ in figure \ref{fig:clkk_pk_rel_error}

\begin{figure}[htb]
    \centering % <-- added
\begin{subfigure}{0.5\textwidth}
  \includegraphics[width=\linewidth]{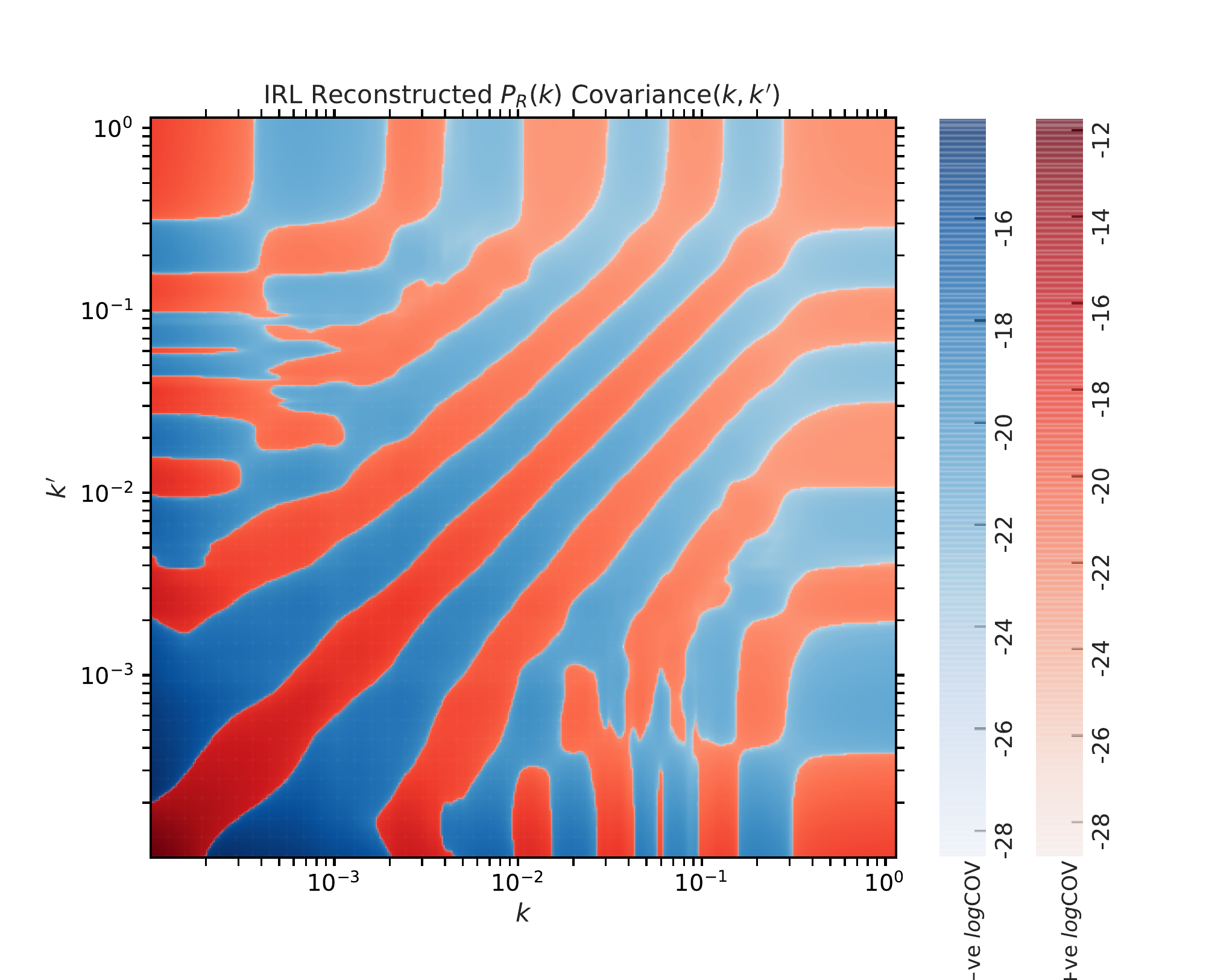}
  \caption{}
  \label{fig:cov}
\end{subfigure}\hfil % <-- added
\begin{subfigure}{0.5\textwidth}
  \includegraphics[width=\linewidth]{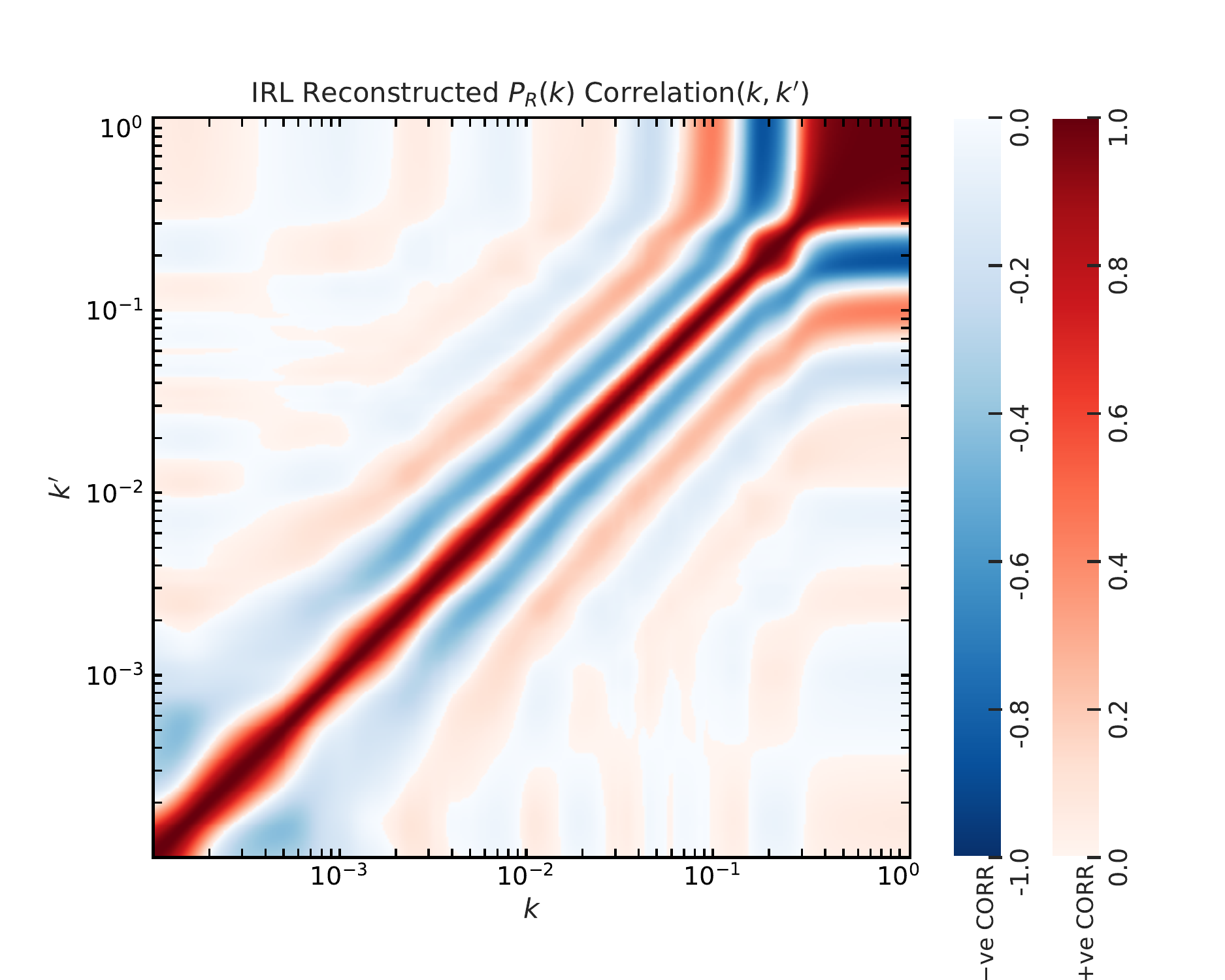}
  \caption{}
  \label{fig:corr}
\end{subfigure}
\caption{ \protect\subref{fig:cov}) is a plot of the $\Sigma_{kk'}$ from the reconstructed $P_{R}(k)$. Reds and Blues denote $\pm$ $\log_{10}\Sigma_{kk'}$ respectively. The plot \protect\subref{fig:corr}) plots the correlations matrix $\rho_{kk'}$ with the Reds, Blues being the $\pm 0\rightarrow 1$ range respectively. }
\label{fig:cov_corr_allk_unbin}
\end{figure}

\begin{figure}[htb]
    \centering % <-- added
\begin{subfigure}{0.55\textwidth}
  \includegraphics[width=\linewidth]{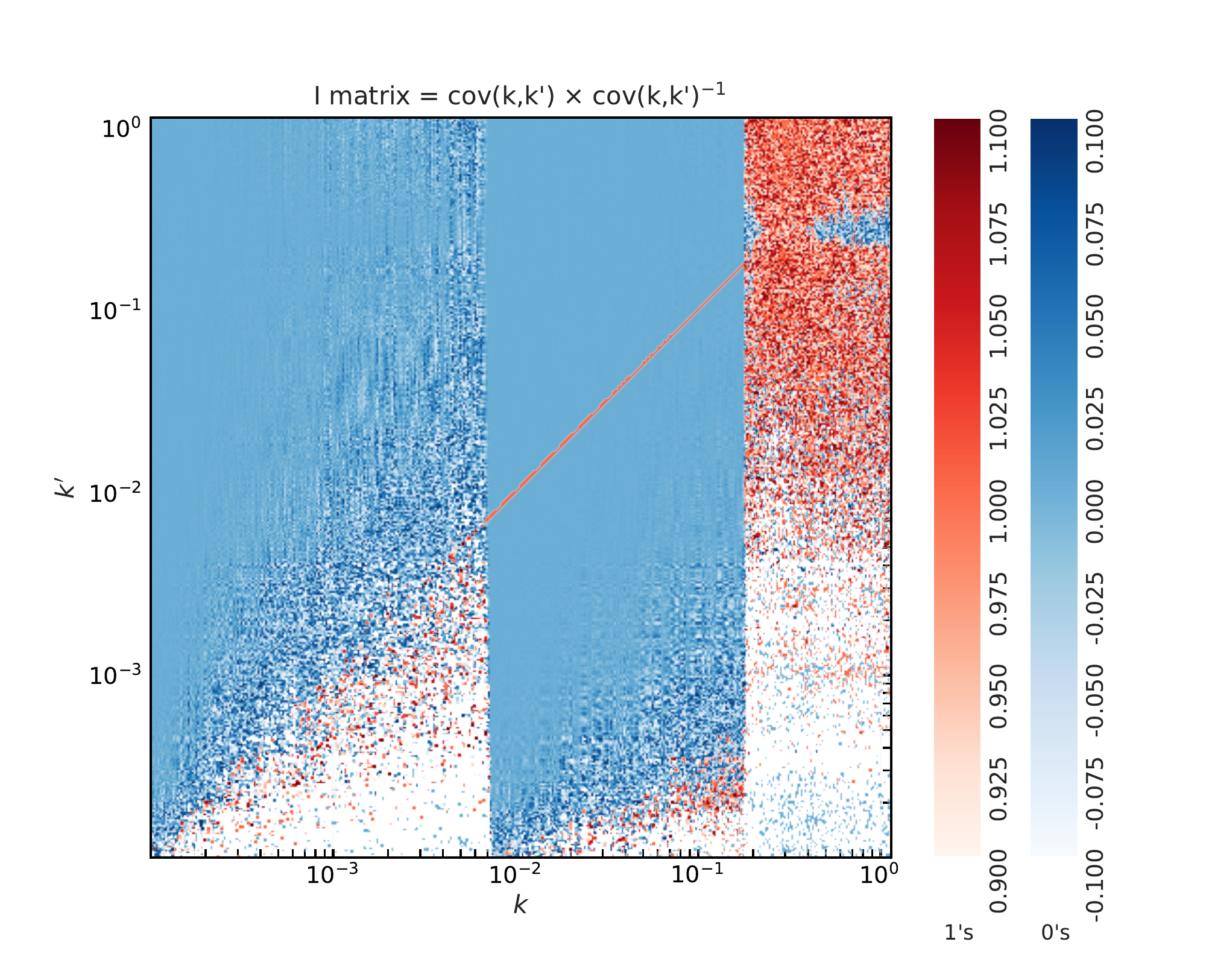}
  \caption{}
  \label{fig:1_0}
\end{subfigure}\hfil % <-- added
\begin{subfigure}{0.45\textwidth}
  \includegraphics[width=\linewidth]{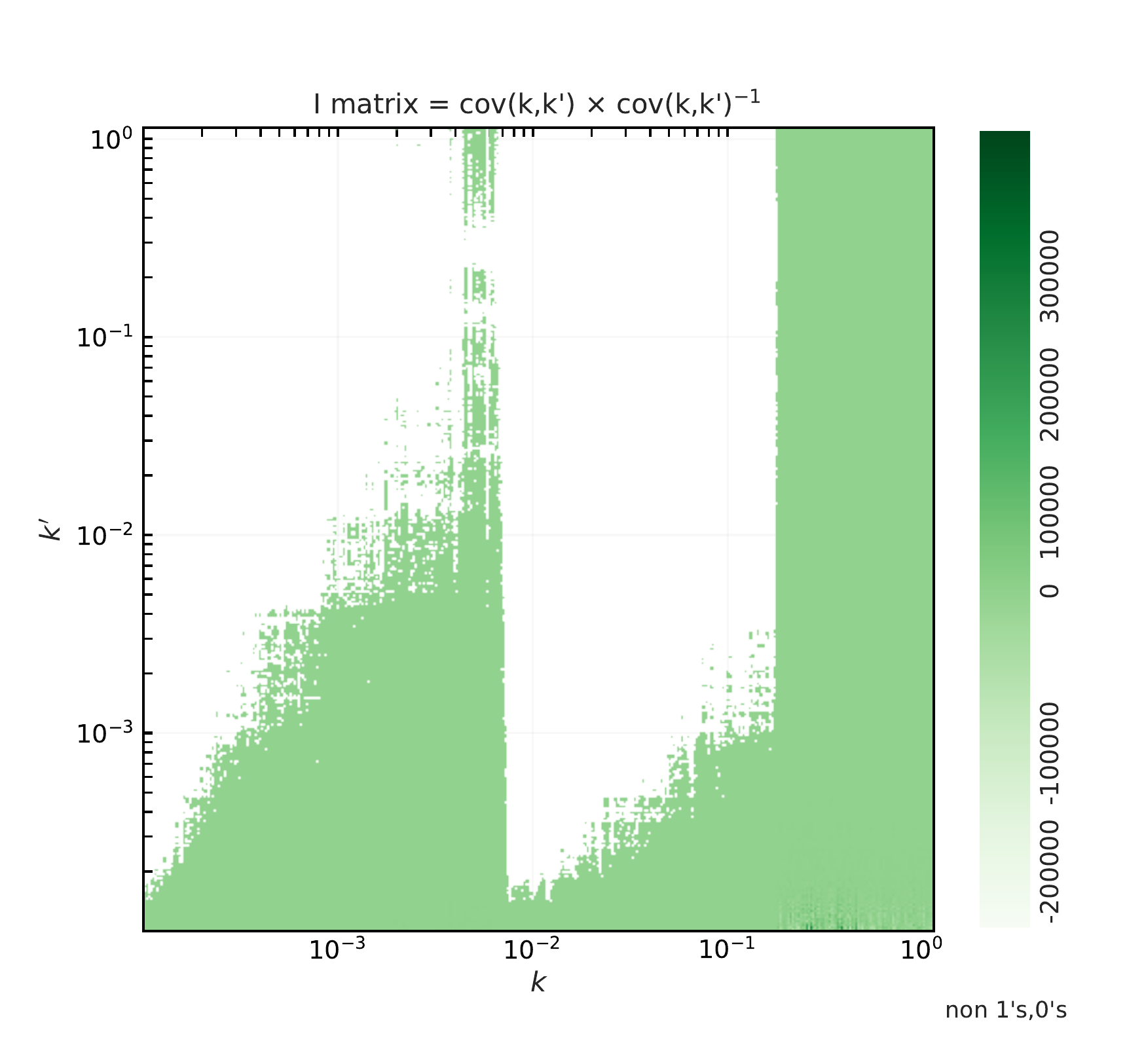}
  \caption{}
  \label{fig:non_1_0}
\end{subfigure}
\caption{ \protect\subref{fig:1_0}) Plots the $\Sigma_{kk'}\times \Sigma_{kk'}^{-1}$ with Reds, Blues being $1\pm0.1$, $0\pm0.1$ respectively . The plot \protect\subref{fig:corr}) shows the same matrix but for the non $1\pm0.1$,$0\pm0.1$ terms.}
\label{fig:idmat_1_0_non}
\end{figure}

We demonstrate the discussion over the sparsity of $P_{R}(k)$ sampling at low $k$, and the highly correlated nature, numerical inaccuracy at high $k$, with the following plots on the covariance matrix $\Sigma_{kk'}$ and the Identity matrix expected from multiplying with its inverse, in figures \ref{fig:cov} and \ref{fig:idmat_1_0_non}.

In figure \ref{fig:cov}, we have plotted the covariance matrix $\Sigma_{kk'}$ from the reconstructed $P_{R}(k)$ as a heatmap with the reds and blues being the positive and negative parts of $\log_{10}\Sigma_{kk'}$. From the plot it is clear that there is correlation between the different free form $P_{R}(k)$ coefficients. This points to our previous discussion on oversampling over $k$ with respect to the number of $L$ data points. 
To study this better we therefore also plot the correlation matrix in \ref{fig:corr}, given by correlation coefficient  \ref{eq:corr_form}. As a result the reds and blues range from $-1$ to $1$ and provide a clearer picture of the correlation.
Again, as expected from our previous discussion of reconstruction over different $k$ ranges, the degree of correlation is very high at high $K$ and fairly strong at low $k$s as well. This opens up room for us to optimize the $k$ sampling comprehensively, and we will demonstrate this using the inverse of $\Sigma_{kk'}$ and the corresponding Identity matrix when we take their product. 

\begin{equation}
\begin{split}
\rho_{kk'} = \frac{\Sigma_{kk'}}{\sigma_{k}\sigma_{k'}}
\end{split}
\label{eq:corr_form}
\end{equation}

We plot the visualisation used for optimizing the $P_{R}(k)$ sampling, in \ref{fig:1_0}, where the first plot shows the $0$s in blue and $1$s in red, within a numerical tolerance range of $\pm0.1 $. Ideally, $\Sigma_{kk'}\cdot\Sigma^{-1}_{kk'}$ should be an Identity matrix with a clear red diagonal of $1$s and blue $0$s in the off diagonals. However it is evident that the reconstruction in the low kernel support regions of $k < 7\times10^{-3}$ and $k > 1.5\times10^{-1}$, is highly correlated and the system of equations being solved by IRL is not well determined. As a result, the second plot \ref{fig:non_1_0}, which should ideally by $0$ throughout, also shows a lot of 'noise' in the regions of low information $k$ reconstruction, as well as the correlated regions. 
Hence using these figures as a guide and the kernel discussion, we can reduce the density of $k$ sampling in those regions and carry out reconstruction based on the number of $L$ data points per $P_{R}(k)$ sample points, such that the system of equations being solved is reasonably well determined and the reconstruction $\Sigma_{kk'}$ is invertible.

\subsection{Results ii) : Sparse $k$ Density}
\label{sub_sec:sprsksim}

In this section we propose an optimization method to reduce the degeneracy of solutions in $P_{R}(k)$ space. In the previous example we are oversampling leading to both correlations over different $k$ as well as degenerate solutions which can all fit the data used. This is expressed by the poorly reconstructed $\Sigma_{kk'}$ matrix, which while numerically invertible, a dot product with its own inverse $\Sigma_{kk'}^{-1}$ does not result in a clean Identity matrix $\bf{I}$. 
We therefore use a novel algorithm where we divide the $k$ sampling grid into multiple subsets and ensure that the number of $P_{R}(k)$ samples in a given subset is less than the number of new $L$ modes that cross a minimum threshold of $0.001\% G_{L}^{\phi\phi}(k)_{max}$.

Other numerical parameters remain the same, including the input data based on the Power Law model with non correlated Cosmic-Variance limited errors bars given in Figure \ref{fig:input_clkk_data}. 

\begin{figure}[htb]
    \centering % <-- added
\begin{subfigure}{0.5\textwidth}
  \includegraphics[width=\linewidth]{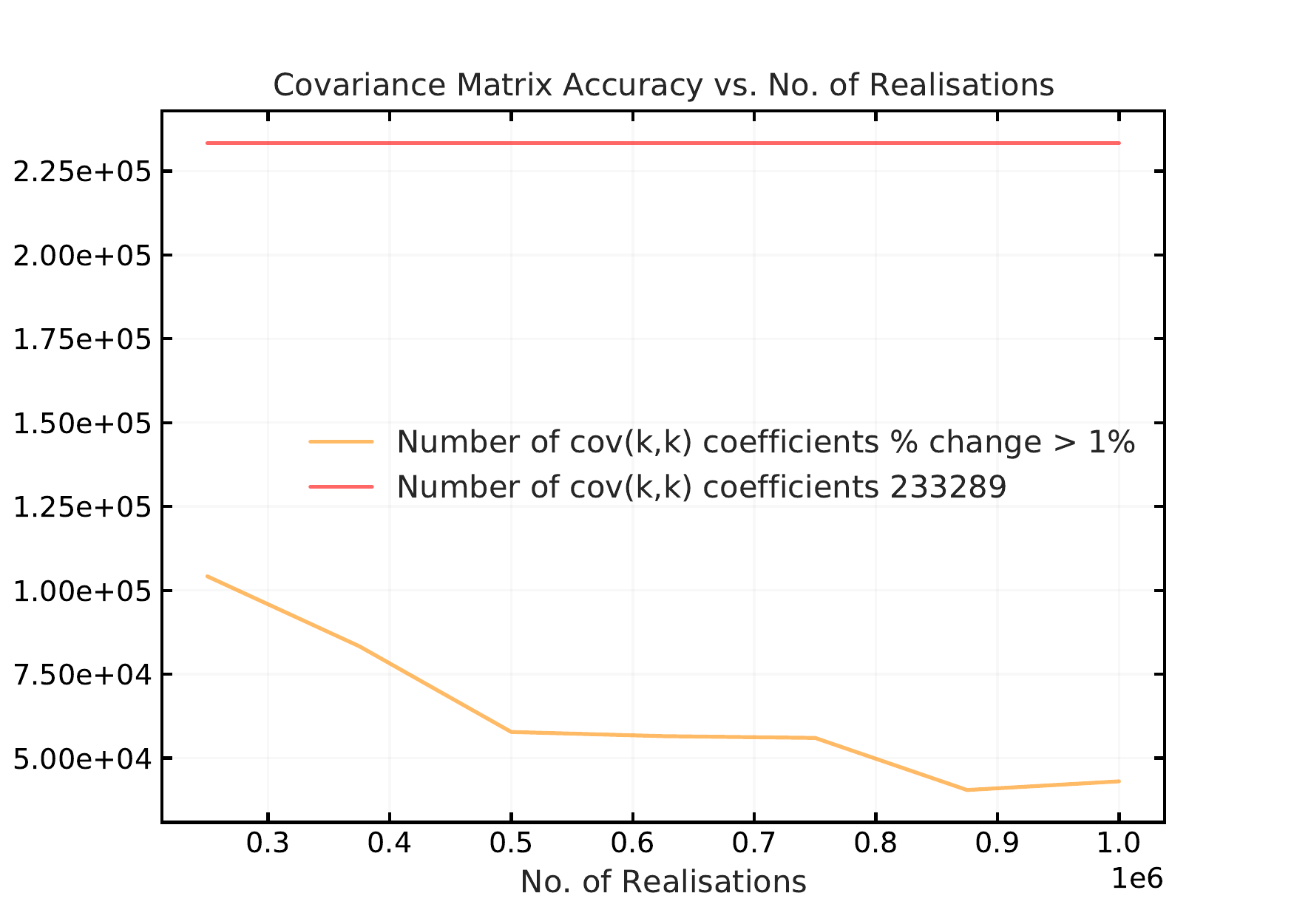}
  \caption{}
  \label{fig:1perc_sat}
\end{subfigure}\hfil % <-- added
\begin{subfigure}{0.5\textwidth}
  \includegraphics[width=\linewidth]{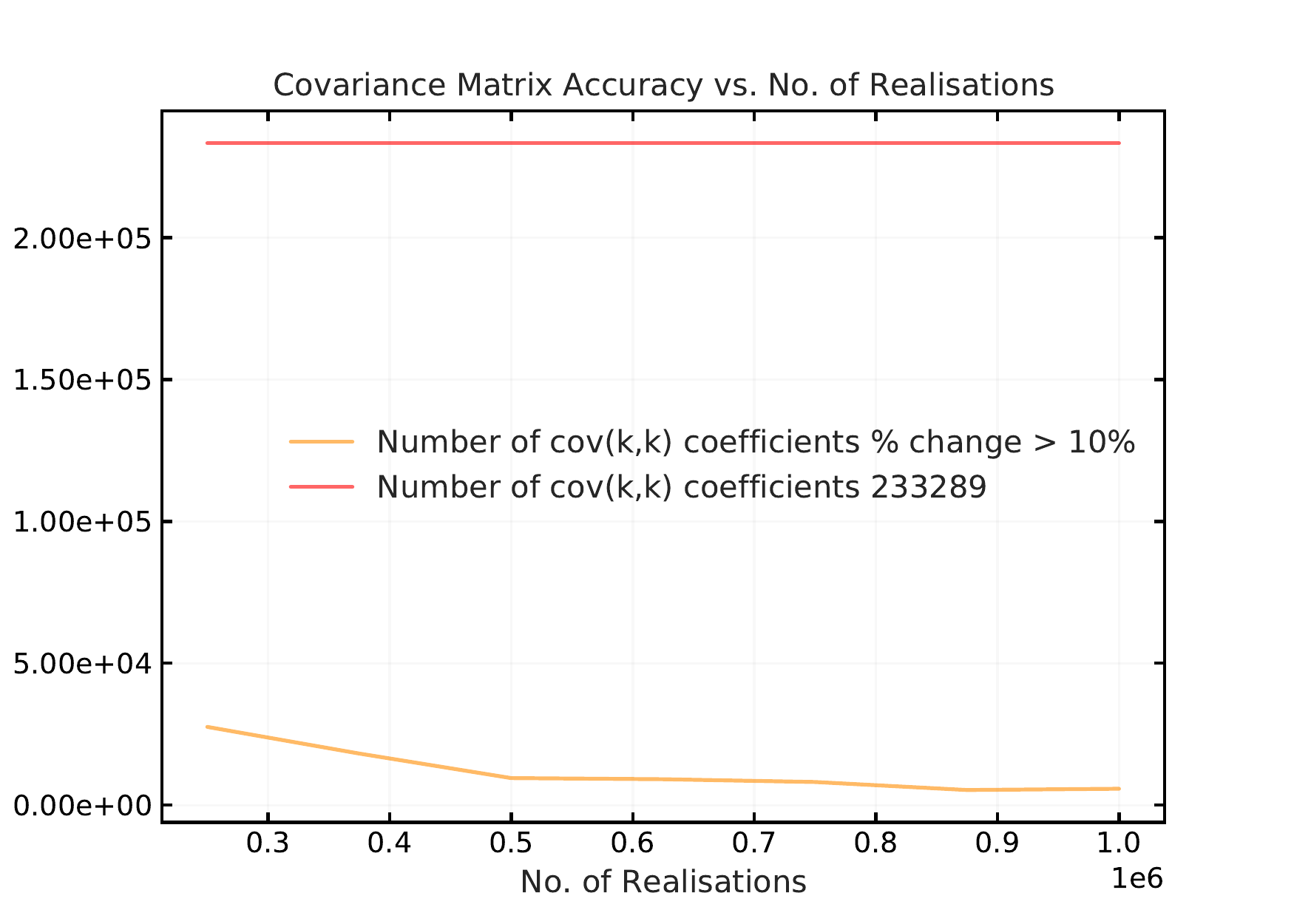}
  \caption{}
  \label{fig:10perc_sat}
\end{subfigure}
\caption{ \protect\subref{fig:1perc_sat}) Plots the number of $\Sigma_{kk'}$ coefficients that exceed a $1\%$ relative error change with respect to the previous realisation set, vs the realisation number set. We see that by \num{2e6} number of realisations, the accuracy saturates. Similarly the plot \protect\subref{fig:1perc_sat}) plots the same but at $10\%$ accuracy threshold, showing similar saturation.}
\label{fig:Feature_DataRlzn_kspars}
\end{figure}

Based on this algorithm we obtain a new $P_{R}(k)$ sampling vector and generate \num{2e6} data samples and IRL reconstruction samples. The cost of computation is lower given lesser number of $P_{R}(k)$. Figure \ref{fig:Feature_DataRlzn_kspars} gives the numerical accuracy of the covariance matrix $\Sigma_{kk'}$ with respect to accuracy again.

\begin{figure}
\centering % \begin{center}/\end{center} takes some additional vertical space
\includegraphics[width=1.00\linewidth]{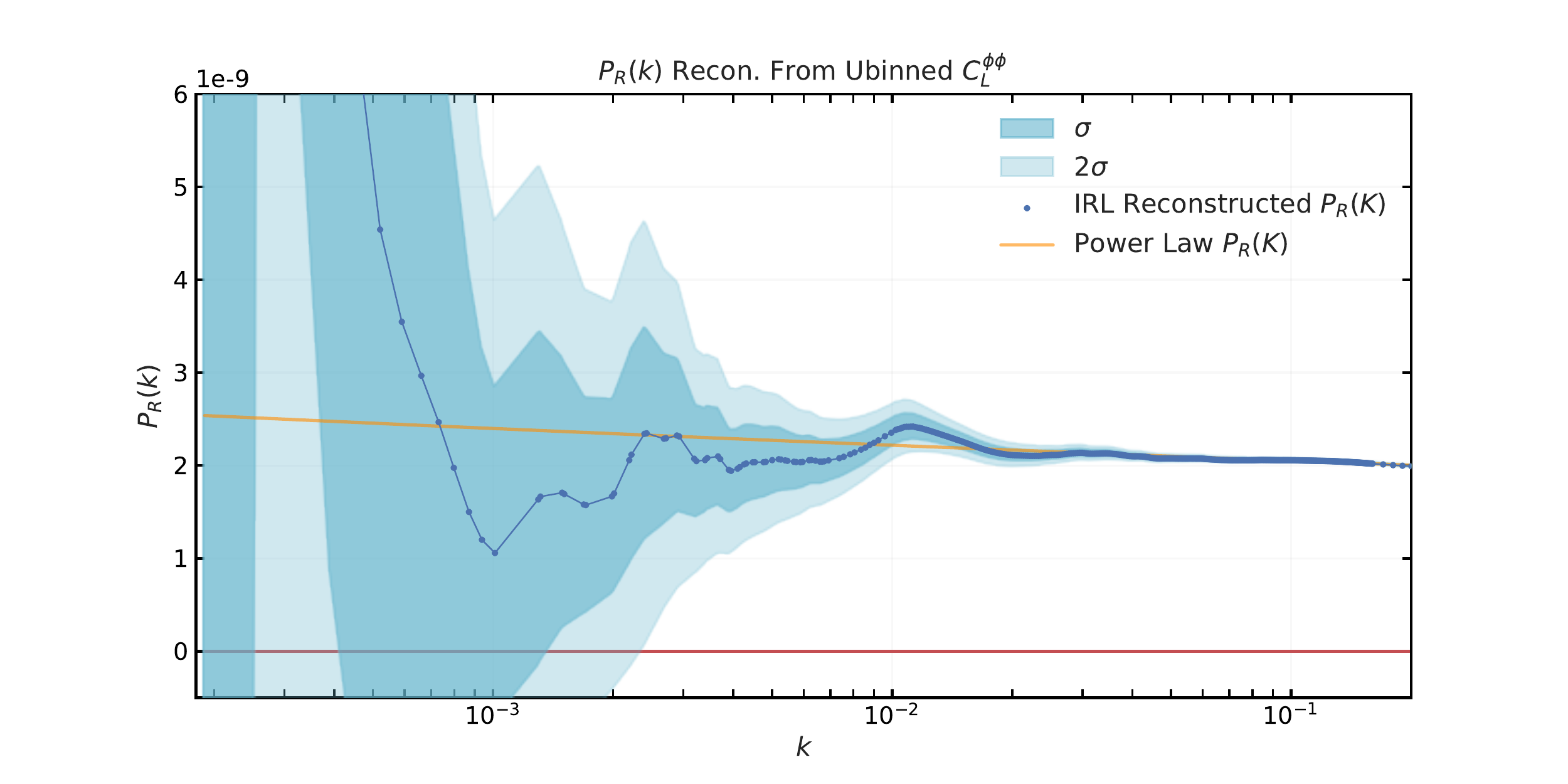}
\caption{Plot of the reconstructed $P_{R}(k)$ (Blue) with $1\sigma,2\sigma$ bands from $2\times10^{6}$ data realisations with cosmic variance error, overplotted on the fiducial Power Law $P_{R}(k)$ (Orange).}
\label{fig:recon_unbin_sprsk_pk_sigband}
\end{figure}

We again plot the reconstructed $P_{R}(k)$ with  $1\sigma, 2\sigma$ error bands in figure \ref{fig:recon_unbin_sprsk_pk_sigband}. , and we impose the constraint by the initial-guess sensitivity information and use the $k$ range given in table \ref{table:sparse_k_unbin_IRL_powlaw}.

\begin{table}
%\emph
{\begin{center}
%s\noindent
\begin{tabular}{ l r r l }
  \toprule
  Parameter & \multicolumn{1}{c}{Values} \\
  \midrule
  Input Data & $C_L^{\kappa\kappa}$ \\
  Error Bars & Cosmic Variance \\
  Input $P_{R}(k)$ Model & $A_s(k/k_*)^{(n_s-1)}$ \\
  $k$ Range & [ \num{e-4} $\rightarrow$ \num{0.2} ] \\
  $N_k$ & 483 \\
  $L$ Range & [ \num{2} $\rightarrow$ \num{2500} ]  \\
  $L$ Binning & Unbinned \\
  Data Realisations & \num{2e6} \\
  \bottomrule
\end{tabular}
\end{center}}
\caption{The simulation parameters over which the IRL reconstruction is carried out.}
\label{table:sparse_k_unbin_IRL_powlaw}
\end{table}

\begin{figure}
\centering % \begin{center}/\end{center} takes some additional vertical space
\includegraphics[width=1.00\linewidth]{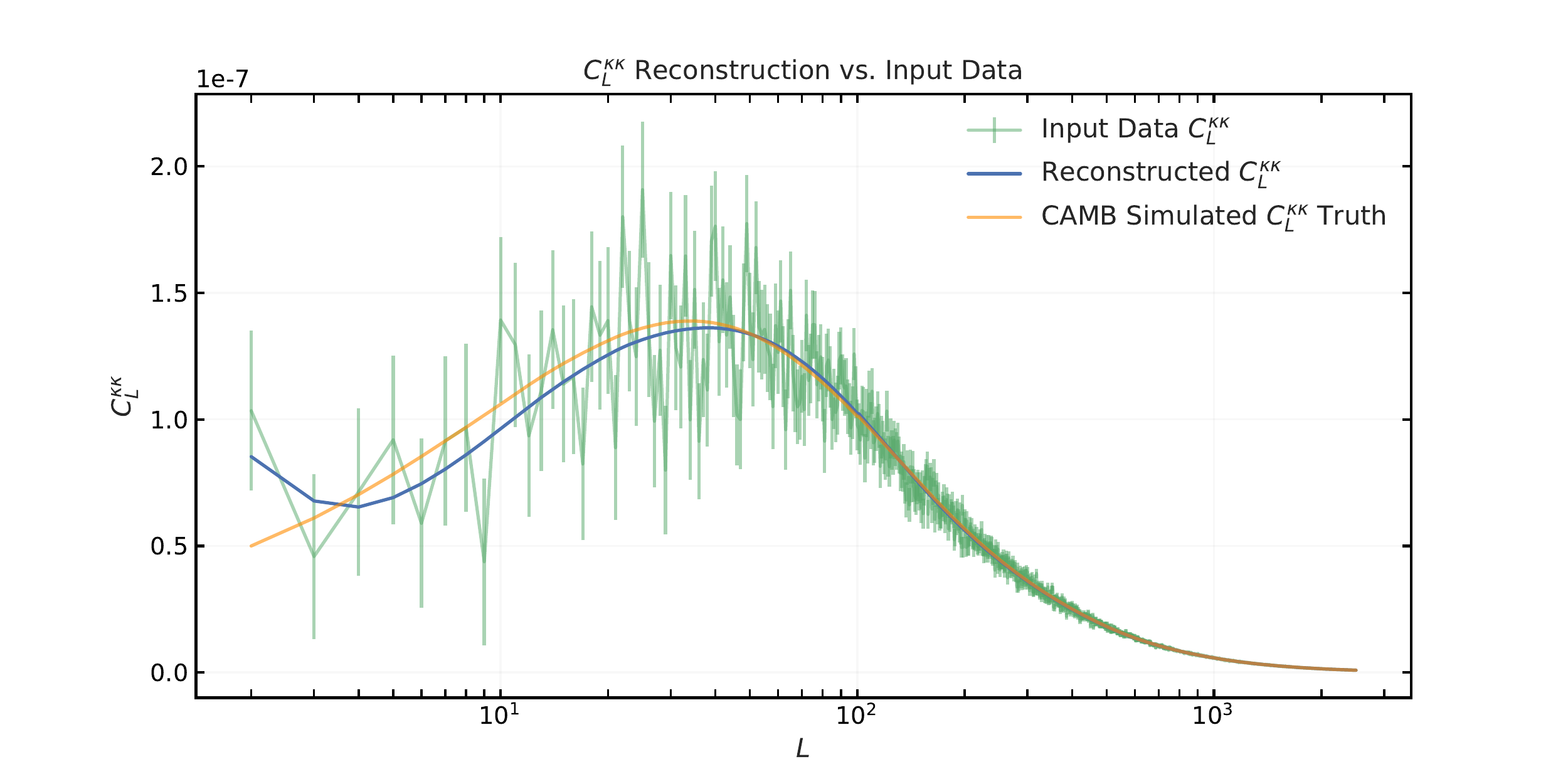}
\caption{Reconstructed $C_L^{\kappa\kappa}$ in Blue plotted over the fiducial CAMB $C_L^{\kappa\kappa}$ in Orange and the input data realisation in Green.}
\label{fig:recon_unbin_sprsk_clkk_vs_inp}
\end{figure}

\begin{figure}
\centering % \begin{center}/\end{center} takes some additional vertical space
\includegraphics[width=1.00\linewidth]{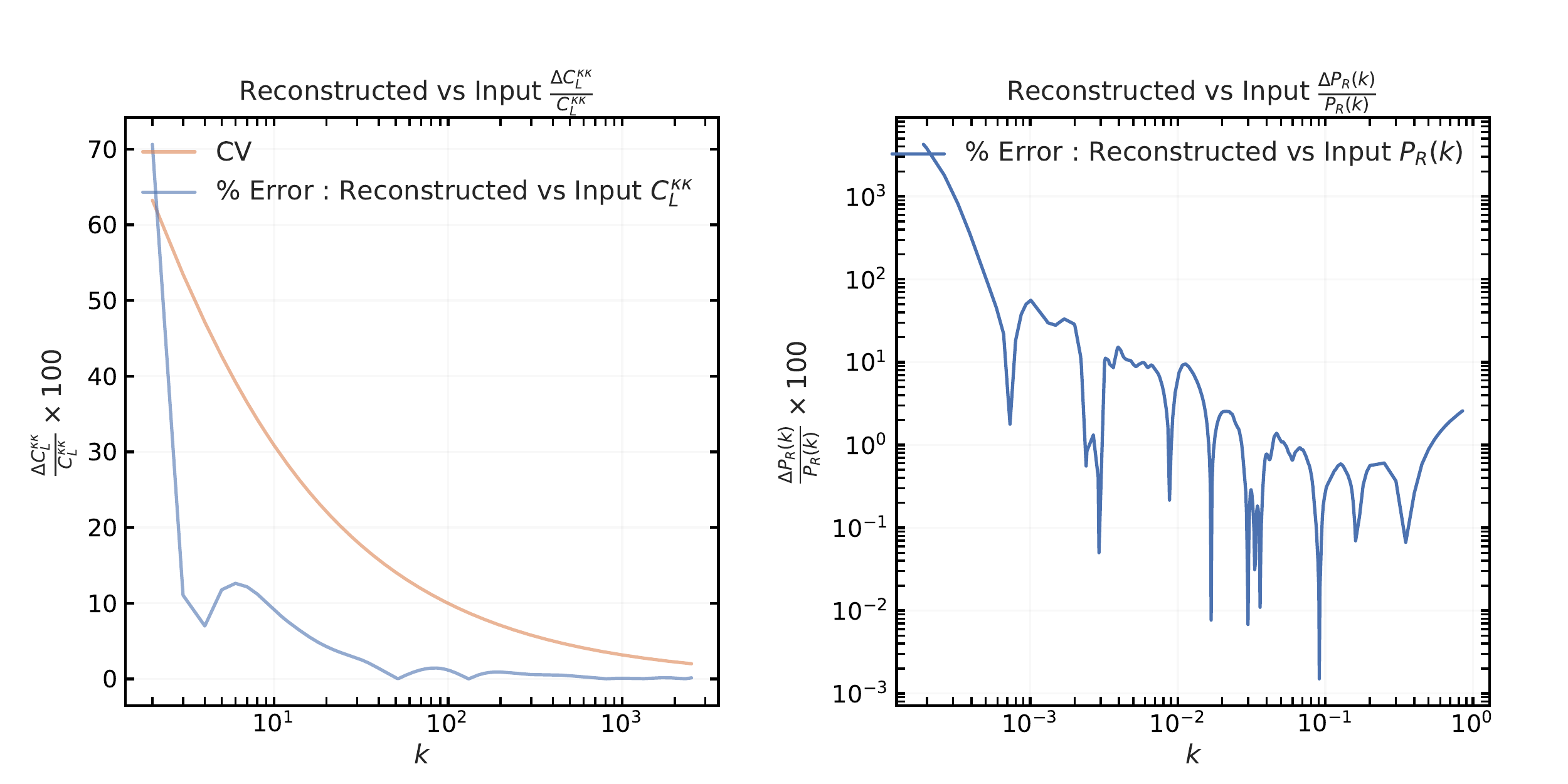}
\caption{The two figures show the relative \% error between the reconstructed $C_L^{\kappa\kappa}$ vs data realisation $C_L^{\kappa\kappa}$, and the reconstructed $P_{R}(k)$ vs input Power Law $P_{R}(k)$}
\label{fig:clkk_pk_rel_error_sprsk}
\end{figure}

The reconstructed $C_L^{\kappa\kappa}$ and its comparison to the input fiducial CAMB simulated $C_L^{\kappa\kappa}$ in in figure \ref{fig:recon_unbin_sprsk_clkk_vs_inp}, and the relative error plots between both reconstructed $C_L^{\kappa\kappa}$ and $P_{R}(k)$ with respect to theoretical CAMB $C_L^{\kappa\kappa}$ and Power Law model in figure \ref{fig:clkk_pk_rel_error_sprsk}, for all $k$ ranges. (We include all $k$ for these plots as they contribute to the high $L$ ranges, however for feature hunting and statistical analysis of $P_{R}(k)$, we will work with the cutoff at $k \leq 0.2 \text{ Mpc}^{-1}$ as defined above in initial guess sensitivity.

\begin{figure}[htb]
    \centering % <-- added
\begin{subfigure}{0.5\textwidth}
  \includegraphics[width=\linewidth]{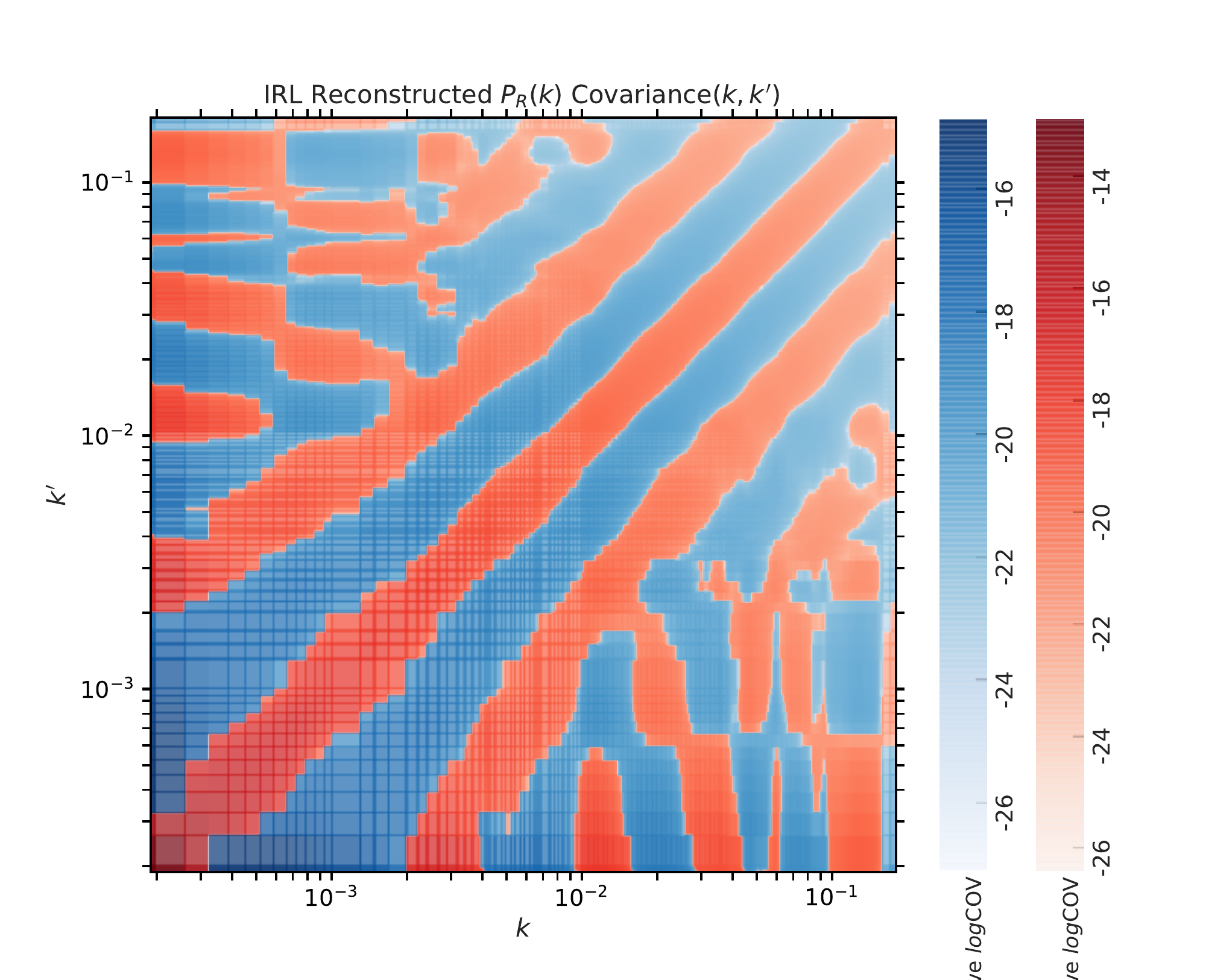}
  \caption{}
  \label{fig:cov_sprsk}
\end{subfigure}\hfil % <-- added
\begin{subfigure}{0.5\textwidth}
  \includegraphics[width=\linewidth]{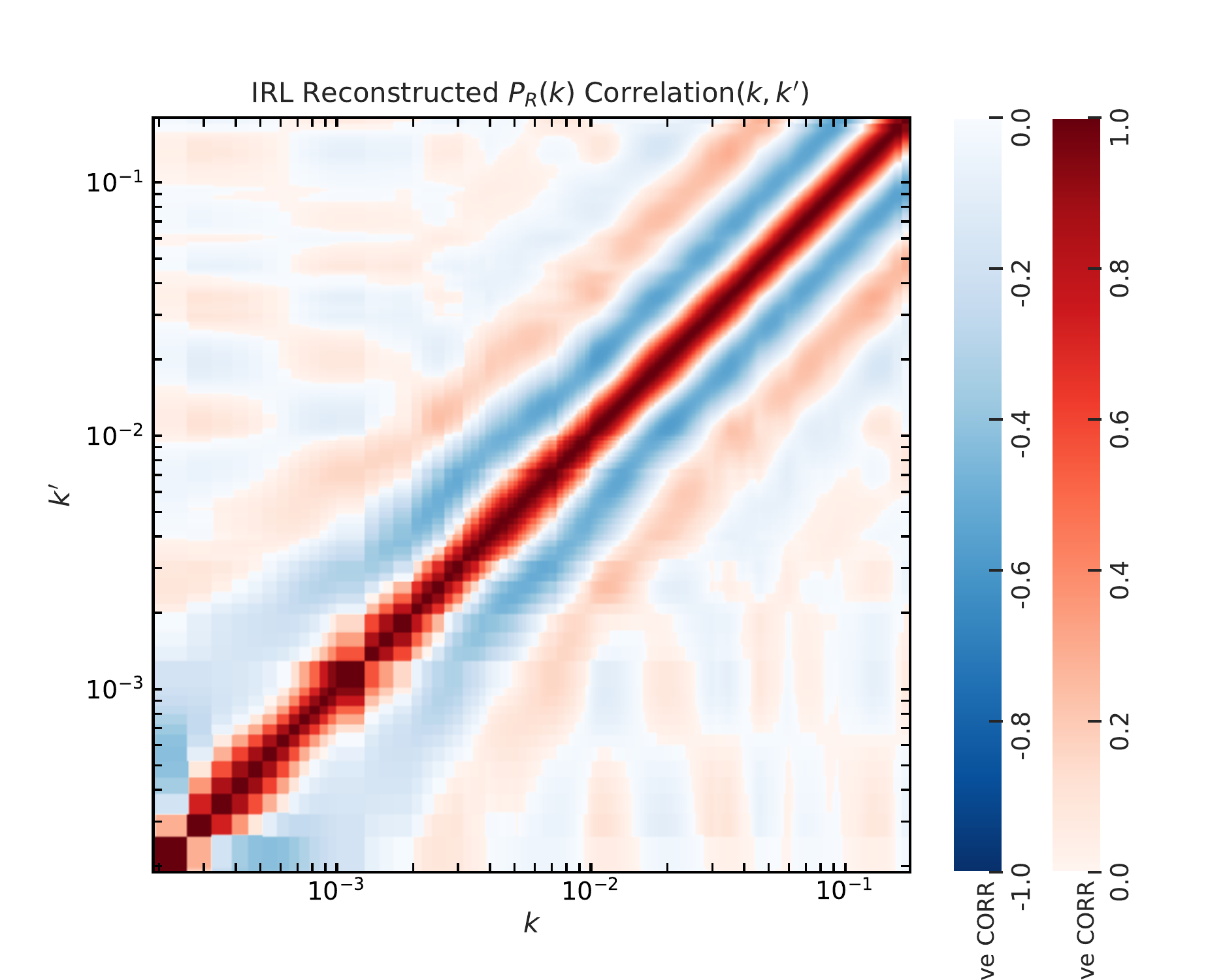}
  \caption{}
  \label{fig:corr_sprsk}
\end{subfigure}
\caption{ \protect\subref{fig:cov_sprsk}) is a plot of the $\Sigma_{kk'}$ from the reconstructed $P_{R}(k)$. Reds and Blues denote $\pm$ $\log_{10}\Sigma_{kk'}$ respectively. Plot \protect\subref{fig:corr_sprsk}) plots the correlations matrix $\rho_{kk'}$ with the Reds, Blues being the $\pm 0\rightarrow 1$ range respectively. }
\label{fig:cov_corr_sprsk_unbin}
\end{figure}

\begin{figure}[htb]
    \centering % <-- added
\begin{subfigure}{0.5\textwidth}
  \includegraphics[width=\linewidth]{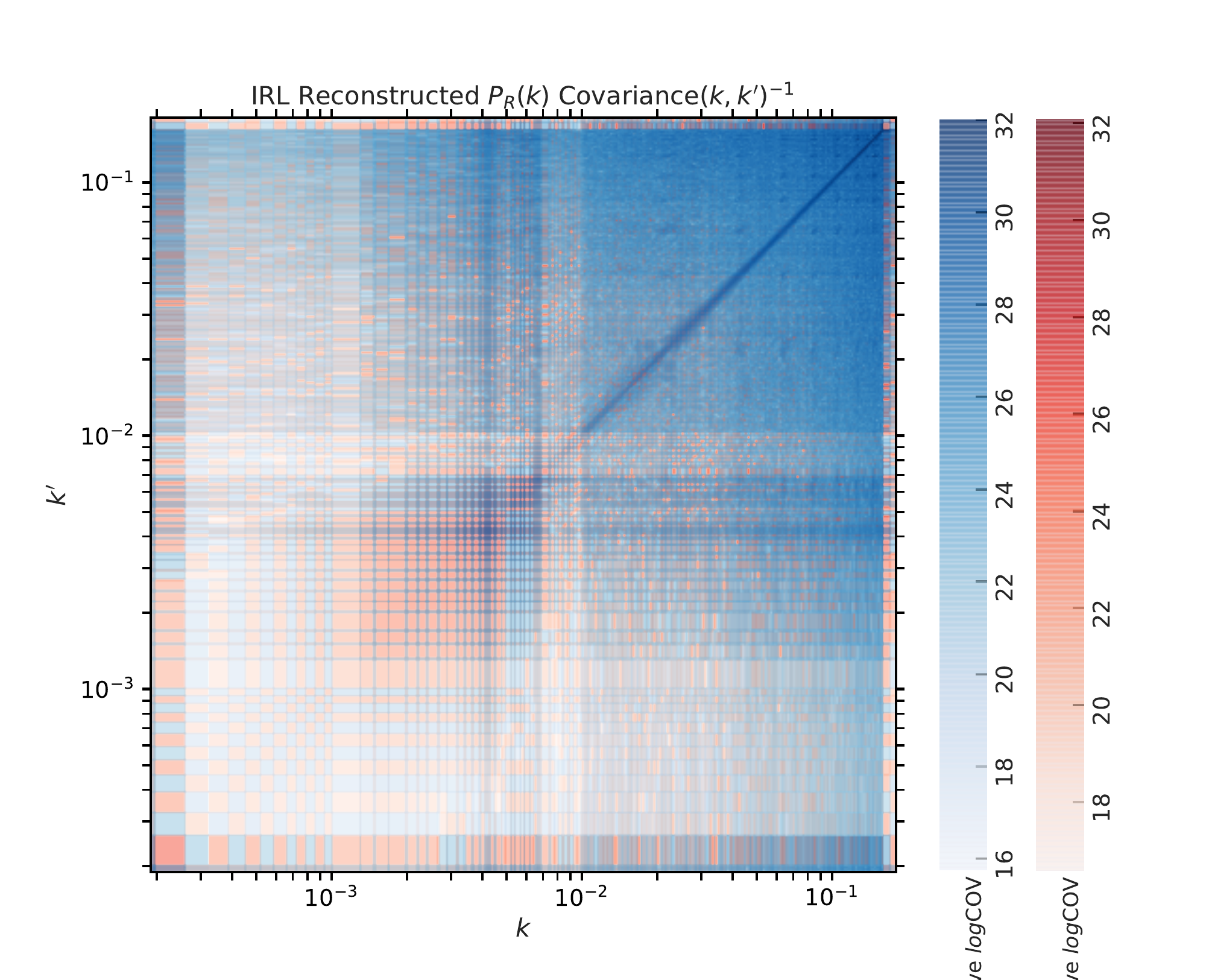}
  \caption{}
  \label{fig:cov_inv}
\end{subfigure}\hfil % <-- added
\begin{subfigure}{0.5\textwidth}
  \includegraphics[width=\linewidth]{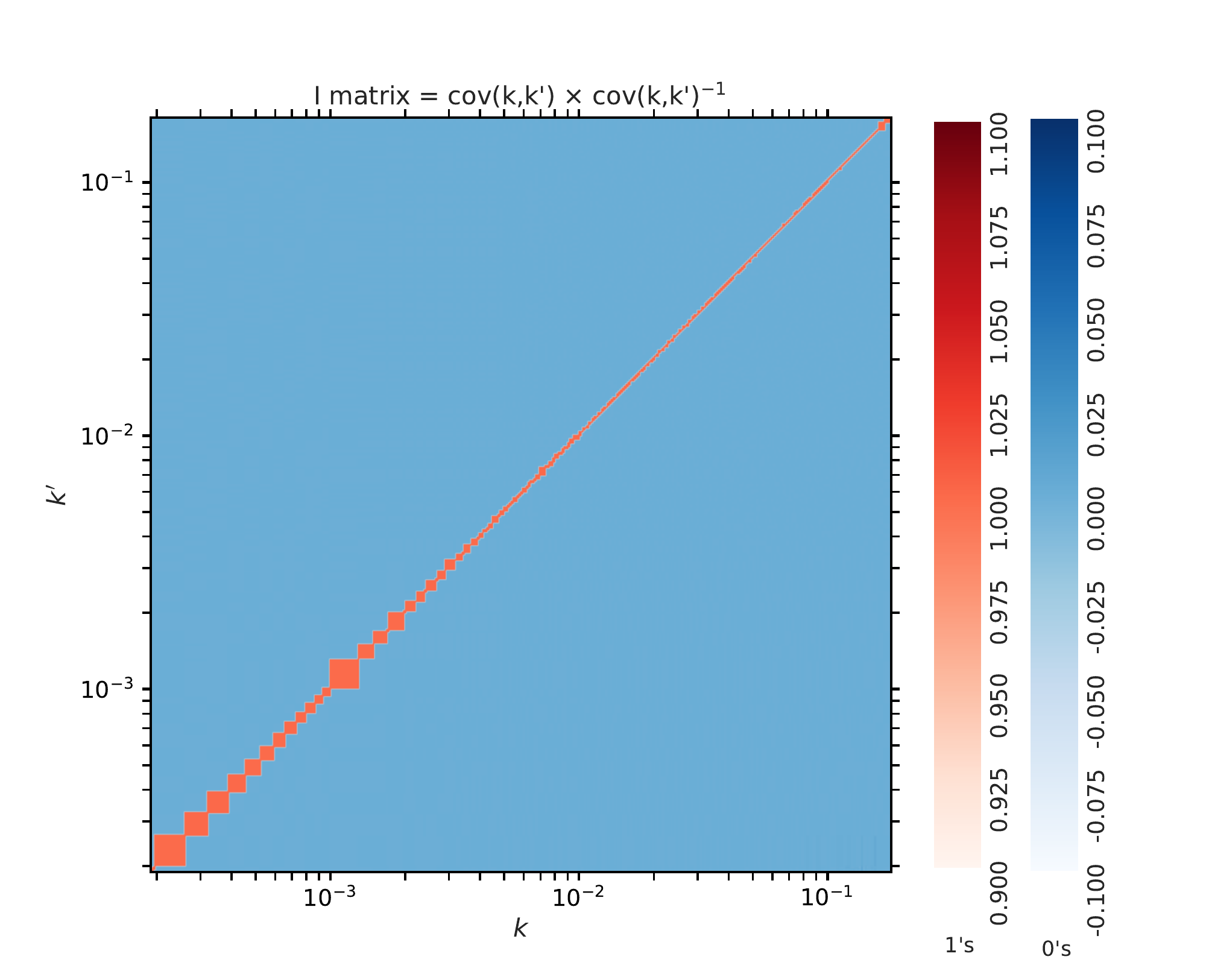}
  \caption{}
  \label{fig:1_0_ksprs}
\end{subfigure}
\caption{ \protect\subref{fig:cov_inv}) Plots the $\Sigma_{kk'}^{-1}$. \protect\subref{fig:1_0_ksprs}) plots the $\Sigma_{kk'}\times \Sigma_{kk'}^{-1}$ with Reds, Blues being $1\pm0.1$, $0\pm0.1$ respectively. A clear Identity matrix is obtained within numerical bounds, showing that the $\Sigma_{kk'}$ is obtained from a unique $P_{R}(k)$ solution.}
\label{fig:idmat_1_0_non_ksprs}
\end{figure}

We should note that we have naively plotted the $1\sigma, 2\sigma$ bands around the reconstructed $P_{R}(k)$, but the standard deviation does not reflect the rest of the properties of the probability distribution of each $P_{R}(k)$ sample. For low $k$ we should also includes the skew statistic, since the $P_{R}(k)$ cannot be negative, being a power spectrum. However for the purpose of this analysis, we will ignore such details as the low $k$ regions have very high reconstruction error even for the ideal cosmic variance error limit in data, and as such are unlikely to ever be precise enough for feature hunting. 
 We also plot the individual $P_{R}(k)$ sample points being evaluated, which are more sparse now, but at the advantage of being statistically more significant overall. This is expressed in the covariance/correlation matrix, its inverse and Identity matrix, $\Sigma_{kk'}$, $\Sigma_{kk'}^{-1}$ and $\bf{I}$. These are plotted in figures \ref{fig:cov_corr_sprsk_unbin} and \ref{fig:idmat_1_0_non_ksprs}

We can see the drastic improvement in the covariance matrix for the new $P_{R}(k)$ reconstruction using the algorithm of $P_{R}(k)$ sampling following the kernel contributions per $L$ above a threshold. From this we can infer that our $P_{R}(k)$ sampling is such that chances of degenerate solutions for $P_{R}(k)$ are reduced. We obtain an invertible $\Sigma_{kk'}$ within numerical limitations. While correlations are still clearly present in \ref{fig:corr_sprsk}, we can better quantify our reconstruction statistically and make a statement if there are significant deviations from the null hypothesis of a Power-Law. 

For our reconstruction we calculate the $\chi^{2}$ value using both the diagonal only error bars $\sigma$ as well as the full covariance matrix $\Sigma_{kk'}$.

\begin{equation}
\begin{split}
\chi^{2}_{diag} & = \sum^{N_k}_{i=1} \frac{[ \hat{P}_{R}(k_i) - P_{R}(k_i)_{0} ]^{2}}{\sigma_{k_i}^{2}} \\
\chi^{2}_{full} & = [ \hat{P}_{R}(k) - P_{R}(k)_{0} ] \Sigma_{kk'}^{-1} [ \hat{P}_{R}(k') - P_{R}(k')_{0} ]^{T}
\end{split}
\end{equation}

The results are given in table \ref{table:kspars_chisq}. It is observed that using the full covariance matrix $\Sigma_{kk'}$ gives us a drastic reduction in the $\chi^2$ value of the reconstruction with respect to power law. The reason for this is because of the correlation present between different $P_{R}(k)$ samples which are taken into account in the off diagonal terms and provide a better statistical estimate of the reconstruction. It is evident that a full reconstruction of the error covariance matrix carries a lot of statistical information of the recovered $P_{R}(k)$ when performed under the optimization we have defined earlier. This procedure helps validate any features being found in the reconstruction.
The \textit{p}-value for the full covariance matrix is close to $1$ given the degrees of freedom and numerical limitations, which says that our null hypothesis if the power law is indistinguishable from the reconstruction, as is expected for the test data generated using the power law. For the diagonal error $\sigma$ only, the \textit{p}-value is 0.976, which is also indistinguishable from the null-hypothesis. 
\begin{table}
%\emph
{\begin{center}
%s\noindent
\begin{tabular}{ l r r l }
  \toprule
  $\chi^{2}$($\hat{P}_{R}(k)- P_{R}(k)_{0}$) & \multicolumn{1}{c}{Values} \\
  \midrule
  $\chi^{2}_{diag}$ & \num{409.52} \\
  $\chi^{2}_{full}$ & \num{56.73} \\
  \bottomrule
\end{tabular}
\end{center}}
\caption{The $\chi^{2}$ values for the reconstructed $\hat{P}_{R}(k)$ with respect to a Power-Law model $P_{R}(k)_{0}$ for $\Sigma_{kk'}$ and diagonal only components.}
\label{table:kspars_chisq}
\end{table}

\section{Discussion}
\label{sec:discuss}

In this paper we analyse a hitherto unexplored data source for ${P}_{R}(k)$ reconstruction, using the $C_L^{\kappa\kappa}$ power spectrum.

We use the standard $\Lambda$CDM cosmology model with Planck best-fit parameters to carry out the MRL deconvolution algorithm and reconstruct a free-form ${P}_{R}(k)$ from simulated data under ideal observation conditions, namely limited by cosmic variance. We also establish reconstruction bounds from a detailed analysis of the transport kernel $G_{LK}^{\kappa\kappa}$ and its applicability for this reconstruction.
We find that the reconstruction limits are from the $k$ range of $2\times10^{-4}$ to $2\times10^{-1}$ and the kernel is optimized when seeking broad features over ${P}_{R}(k)$ due to the smooth nature of the kernel $G_{LK}^{\kappa\kappa}$.

We also establish new paradigms of statistical precision for the reconstruction algorithm and provide a prescription for ${P}_{R}(k)$ sampling based on the kernel properties and verify the improvements using the full ${P}_{R}(k)$ reconstruction covariance matrix for $\Sigma_{kk'}$ and show that we can successfully avoid degenerate solutions by this method. These methods can potentially by applied to IRL reconstruction from other power spectra as well.

We also carry out a $\chi^{2}$ of reconstruction vs null-test estimation to demonstrate that the algorithm is well-behaved and reconstructs a ${P}_{R}(k)$, indistinguishable from the input power law model, without introducing statistically significant spurious features. We conclude that the algorithm is robust with respect to our reconstruction goals. We also show that accounting for the full covariance matrix significantly reduces the $\chi^{2}$-value, which is expected given the correlated nature of the reconstructed free form ${P}_{R}(k)$.

We should also mention that reconstruction from $C_L^{\kappa\kappa}$ has an added advantage in that the power spectrum does not contain secondary distortions such as the weak lensing damping of acoustic peaks of $C_L^{TT}$ at high $L$s. Reconstruction from such data currently requires a delensing template to be subtracted from $\tilde{C}_L^{TT}$. Hence $C_L^{\kappa\kappa}$ is a cleaner probe in this aspect and can be used as a starting point for ${P}_{R}(k)$ reconstructions. 

We expect to carry out more work on this estimator and address several key concerns, from improving the statistical inference of the reconstructions, its behaviour on non power law input modelled data, as well as its applicability in improving ${P}_{R}(k)$ reconstruction jointly with existing CMB anisotropy power spectra and lensing corrections. We also expect to study the estimator behaviour on binned data as well as its application on actual experimental data from Planck and future full-sky CMB missions.

\section{Acknowledgements}
\label{sec:acknow}

This work has been carried out with the support of the Council for Science and Industrial Research (CSIR) and University Grants Commission (UGC) graduate funding at Inter-University Centre for Astronomy and Astrophysics (IUCAA), Pune for RSC.  We would like to acknowledge Shabbir Shaikh for many invaluable discussions and key insights into the paper. We also thank Suvodip Mukherjee for his useful input on Planck covariance matrices. The work acknowledges the use of the IUCAA High Performance Computing facility.

\begin{appendices}

\section{Initial Guess : Supplementary Figures}

\begin{figure}[htb]
%%\centering % \begin{center}/\end{center} takes some additional vertical space
\begin{subfigure}{1\textwidth}
\includegraphics[width=1.00\linewidth]{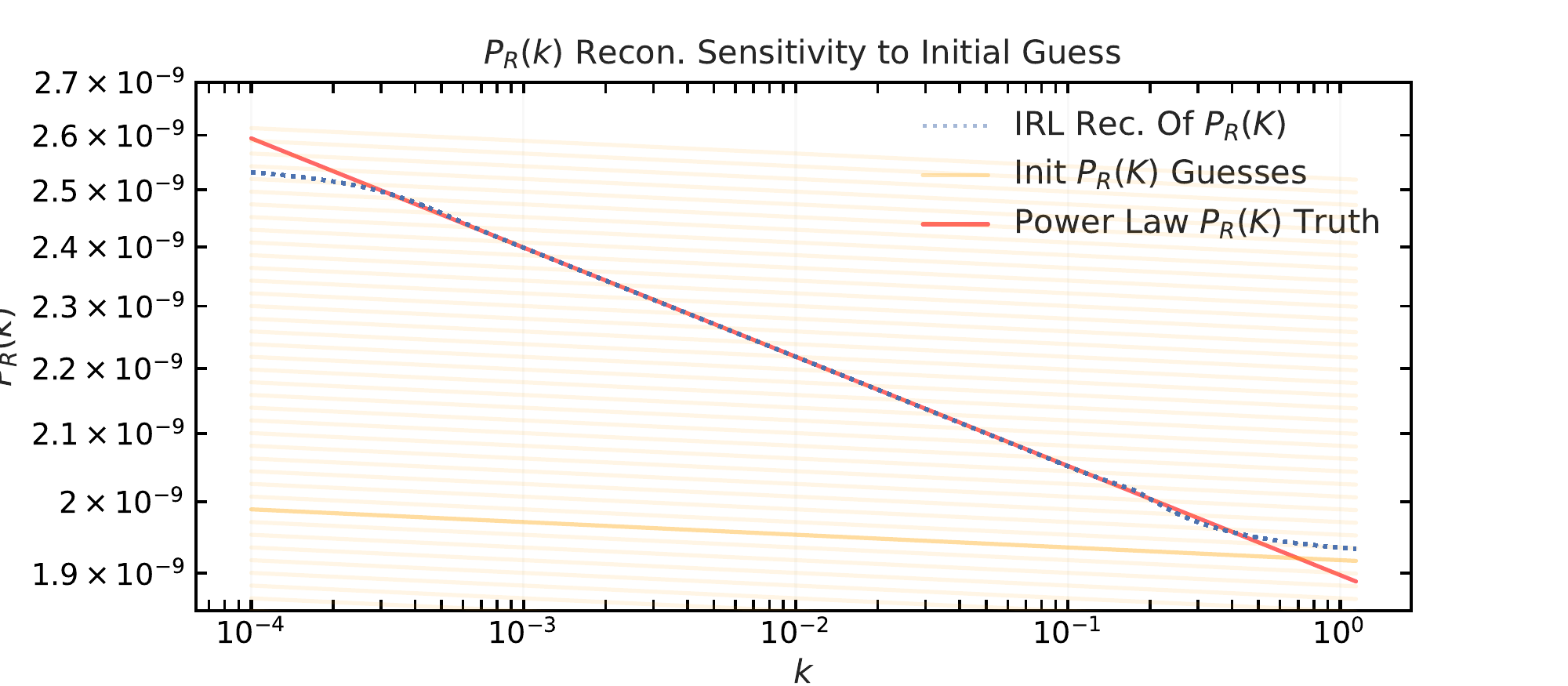}
%% "\includegraphics" is very powerful; the graphicx package is already loaded
%\caption{} 
\label{fig:up_1}
\end{subfigure}
\hspace*{\fill} % separation between the subfigures
\begin{subfigure}{1\textwidth}
\includegraphics[width=1.00\linewidth]{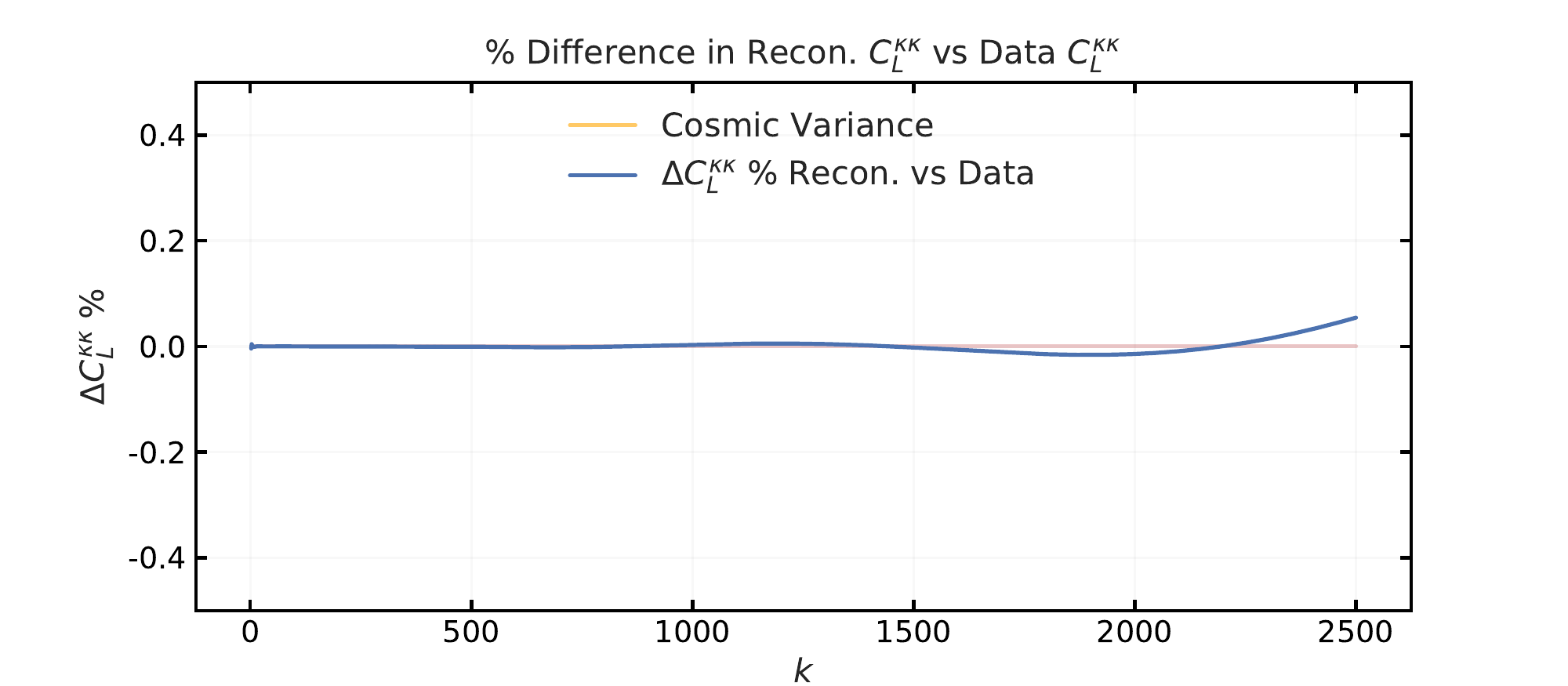}
%% "\includegraphics" is very powerful; the graphicx package is already loaded
%\caption{} 
\label{fig:up_2}
\end{subfigure}
\caption{Figure \protect\subref{fig:slope_1}) shows the reconstructed $P_{R}(K)$ in blue dashed lines, given different initial guesses $P_{R}(K)^{(i=0)}$ in yellow lines varying by intercept $k*$ for a given slope $n_s-0.97$. The red line shows the original injected power spectrum $P_{R}(K)$. Figure \protect\subref{fig:slope_2}) shows the relative \% difference in the reconstructed $\hat{C}_L^{\kappa\kappa}$ and the input data $C_L^{\kappa\kappa}$. It is evident that the slope of the initial guess dominates the reconstruction. }
\label{fig:init_guess_up}
\end{figure}

\begin{figure}[htb]
%%\centering % \begin{center}/\end{center} takes some additional vertical space
\begin{subfigure}{1\textwidth}
\includegraphics[width=1.00\linewidth]{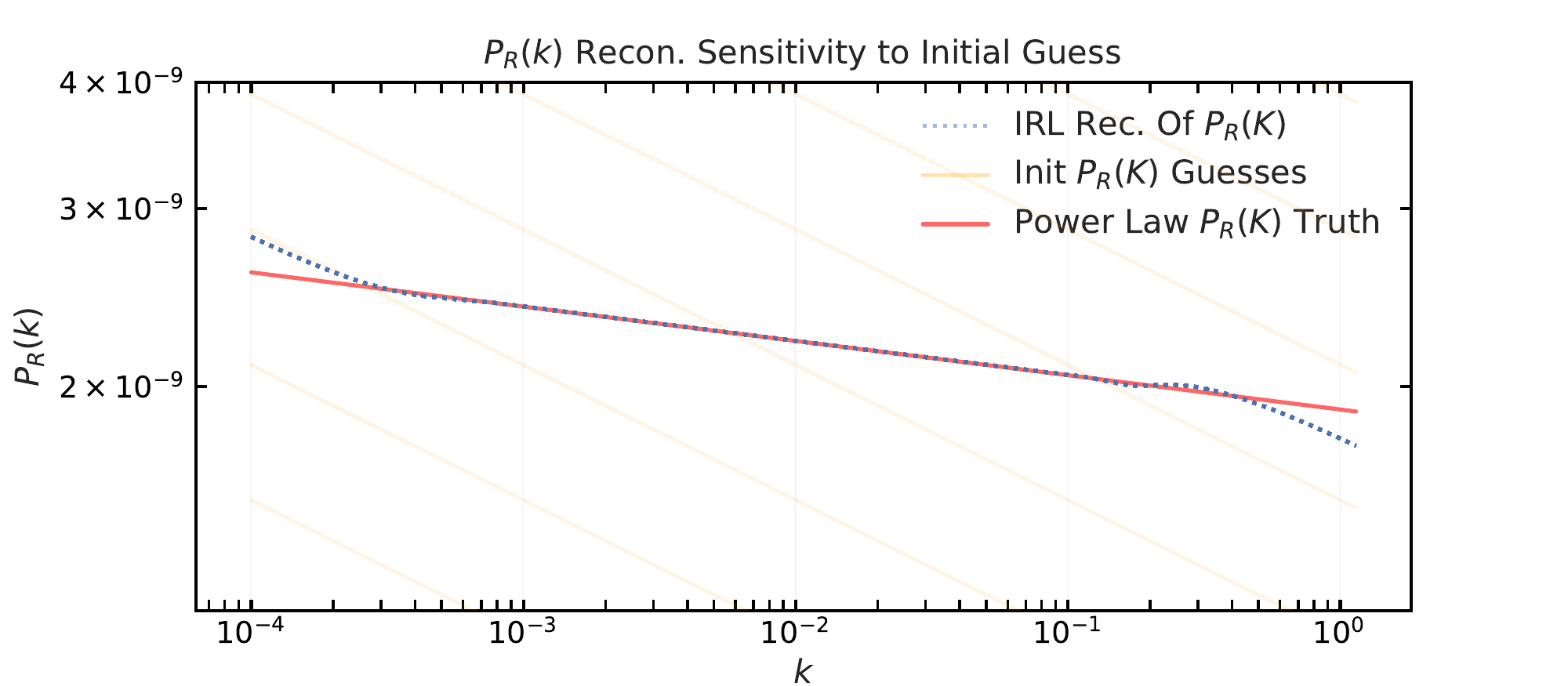}
%% "\includegraphics" is very powerful; the graphicx package is already loaded
%\caption{} 
\label{fig:down_1}
\end{subfigure}
\hspace*{\fill} % separation between the subfigures
\begin{subfigure}{1\textwidth}
\includegraphics[width=1.00\linewidth]{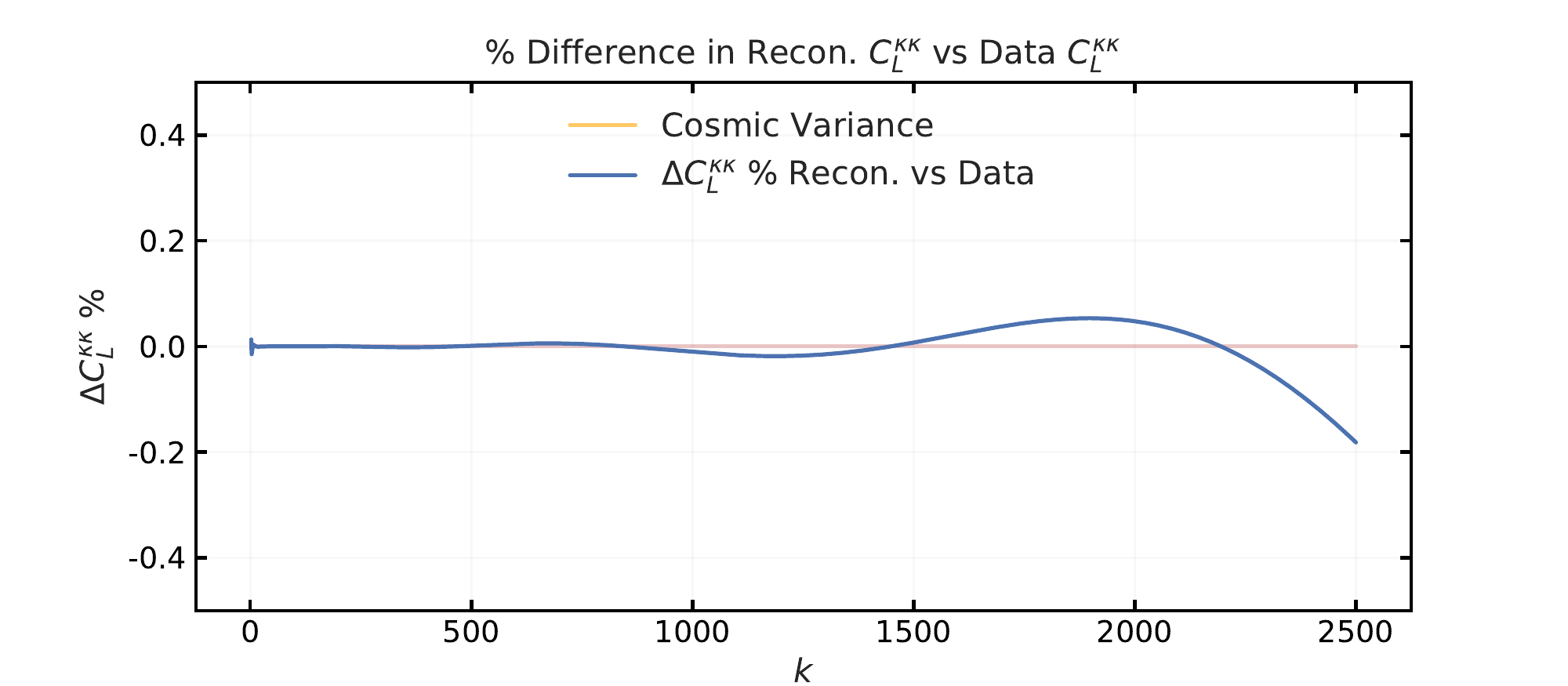}
%% "\includegraphics" is very powerful; the graphicx package is already loaded
%\caption{} 
\label{fig:down_2}
\end{subfigure}
\caption{Figure \protect\subref{fig:slope_1}) shows the reconstructed $P_{R}(K)$ in blue dashed lines, given different initial guesses $P_{R}(K)^{(i=0)}$ in yellow lines varying by intercept $k*$ for a given slope $n_s-1.1$. The red line shows the original injected power spectrum $P_{R}(K)$. Figure \protect\subref{fig:slope_2}) shows the relative \% difference in the reconstructed $\hat{C}_L^{\kappa\kappa}$ and the input data $C_L^{\kappa\kappa}$. It is evident that the slope of the initial guess dominates the reconstruction. }
\label{fig:init_guess_down}
\end{figure}

\end{appendices}

\end{document}